\documentclass[12pt]{article}
\usepackage[english]{babel}
\usepackage[latin1]{inputenc}
\usepackage{color}

\usepackage[intlimits]{amsmath}
\usepackage{amssymb}
\usepackage{slashed}

\usepackage{url}
\usepackage{graphicx}

\numberwithin{equation}{section}

\long\def\symbolfootnote[#1]#2{\begingroup\def\thefootnote{\fnsymbol{footnote}}
\footnote[#1]{#2}\endgroup}

\def\lsim{\mathrel{\raise.3ex\hbox{$<$\kern-.75em\lower1ex\hbox{$\sim$}}}}
\def\gsim{\mathrel{\raise.3ex\hbox{$>$\kern-.75em\lower1ex\hbox{$\sim$}}}}

\newcommand{\fdm}{{\psi_\text{DM}}}
\newcommand{\nn}{N}
\newcommand{\med}{\Sigma}

\newcommand{\lamFL}{\lambda_{\ell\psi}^L} 
\newcommand{\lamFR}{\lambda_{\ell\psi}^R} 
 
\newcommand{\lamNNL}{\lambda_{\ell\nn}^L} 
\newcommand{\lamNNR}{\lambda_{\ell\nn}^R}

\newcommand{\GeV}{\,{\rm GeV}}

\newcommand{\TeV}{\,{\rm TeV}}
\newcommand{\s}{\,{\rm s}}
\renewcommand{\sec}{\s}
\newcommand{\sr}{\,{\rm sr}}
\newcommand{\cm}{\,{\rm cm}}

\newcommand{\Oon}{\Omega_{\text{on}}}
\newcommand{\Ooff}{\Omega_{\text{off}}}
\newcommand{\Non}{N_{\text{on}}}

\newcommand{\Jdm}{J_{\text{dm}}}
\newcommand{\Aeff}{A_{\text{eff}}}

\begin{document}

\setlength{\unitlength}{1mm}

\date{\mbox{ }}

\title{
{\normalsize
MPP 2010-145\hfill\mbox{}\\
TUM-HEP 777/10\hfill\mbox{}\\}
\vspace{1.5cm} 
\bf Gamma-Ray Lines from\\Radiative Dark Matter Decay\\[8mm]}

\author{Mathias Garny$^a$, Alejandro Ibarra$^a$, David Tran$^a$, Christoph
Weniger$^b$\\[2mm]
{\normalsize\it a  Physik-Department T30d, Technische Universit\"at
M\"unchen,}\\[-0.05cm]
{\it\normalsize James-Franck-Stra\ss{}e, 85748 Garching, Germany}\\[2mm]
{\normalsize\it b Max-Planck-Institut f\"ur Physik, M\"unchen}\\[-0.05cm]
{\it\normalsize F\"ohringer Ring 6, 80805 M\"unchen, Germany}}
\maketitle

\thispagestyle{empty}

\begin{abstract}
\noindent
The decay of dark matter particles which are coupled predominantly to charged
leptons has been proposed as a possible origin of excess high-energy positrons
and electrons observed by cosmic-ray telescopes PAMELA and Fermi LAT. Even
though the dark matter itself is electrically neutral, the tree-level decay of
dark matter into charged lepton pairs will generically induce radiative
two-body decays of dark matter at the quantum level. Using an effective theory
of leptophilic dark matter decay, we calculate the rates of radiative two-body
decays for scalar and fermionic dark matter particles. Due to the absence of
astrophysical sources of monochromatic gamma rays, the observation of a line
in the diffuse gamma-ray spectrum would constitute a strong indication of a
particle physics origin of these photons. We estimate the intensity of the
gamma-ray line that may be present in the energy range of a few TeV if the
dark matter decay interpretation of the leptonic cosmic-ray anomalies is
correct and comment on observational prospects of present and future Imaging
Cherenkov Telescopes, in particular the CTA.
\end{abstract}

\newpage

\section{Introduction}
The existence of dark matter is now established beyond reasonable doubt by a
variety of independent observations~\cite{Bertone:2004pz}. These require the
presence of substantial amounts of non-baryonic dark matter at vastly
different scales ranging from individual galaxies to superclusters and
filaments. Despite the overwhelming amount of gravitational evidence, however,
no unambiguous evidence for non-gravitational dark matter interactions has
been discovered to this day.  Since a determination of the particle nature of
the dark matter from its gravitational interactions alone is impossible,
searches for non-gravitational signatures of dark matter are of paramount
importance.

One of the principal approaches to its identification is the indirect
detection of dark matter via searches for exotic components in the cosmic
radiation produced by dark matter interactions with Standard Model particles.
For weakly interacting massive particles (WIMPs), it was pointed out some
decades ago that the dark matter self-annihilation processes that can yield
the correct thermal relic abundance might still occur today at a rate high
enough to give rise to a flux of cosmic rays and photons that may be
detectable by suitable telescopes. However, the self-annihilation of dark
matter is not the only possible scenario for indirect dark matter detection.
Namely, the dark matter might be unstable and decay into Standard Model
particles, even though it must have a very long lifetime in order to survive
from its production in the early Universe to the present day. If the dark
matter is not perfectly stabilized by some unbroken symmetry, however, the
possiblity exists that its decay products may leave visible traces in
cosmic-ray fluxes. Indeed, in some well-motivated models the dark matter
particles are not perfectly stable, but decay with cosmological lifetimes
(see, e.g., \cite{Eichler:1989br, Buchmuller:2007ui, Nardi:2008ix,
Arvanitaki:2008hq, Ibarra:2009bm, Hamaguchi:2008ta, Chen:2008yi,
Ibarra:2008kn, Pospelov:2008rn, Boyarsky:2009ix, Arina:2009uq,
Carone:2010ha}).

Among the possible indirect detection channels, measurements of cosmic-ray
antimatter are particularly sensitive to exotic contributions from dark
matter. Interestingly, the PAMELA telescope has recently confirmed the
existence of a dramatic rise in the positron fraction extending up to energies
of at least 100 GeV~\cite{Adriani:2008zr}, in stark contrast with expectations
from conventional models of cosmic-ray production and
propagation~\cite{Moskalenko:1997gh}. Furthermore, the Fermi Gamma-ray Space
Telescope has observed a total flux of electrons and positrons that is harder
than expected~\cite{Abdo:2009zk, Ackermann:2010ij}, also indicating the
possible presence of an additional component of charged cosmic-ray
leptons~\cite{Grasso:2009ma}.

Various astrophysical explanations for these unexpected behaviors have been
proposed~\cite{Hooper:2008kg,Blasi:2009hv,Ahlers:2009ae}. Arguably more
exciting, however, is the possibility that the excess of positrons and
electrons is due to the annihilation or decay of dark matter particles.
However, measurements of cosmic-ray antiprotons, in particular measurements of
the antiproton-to-proton ratio by PAMELA~\cite{Adriani:2008zq,Adriani:2010rc},
yield stringent constraints on the fraction of dark matter decays or
annihilations into hadronic final states. This has lead some authors to
consider `leptophilic' models of dark matter~\cite{Ibarra:2009bm, Fox:2008kb,
Kyae:2009jt, Bi:2009uj, Davoudiasl:2009dg, Spolyar:2009kx, Cohen:2009fz,
Chun:2009zx, Haba:2010ag, Kopp:2009et}, where the dark matter is coupled
predominantly or exclusively to charged leptons.  In the following, we
consider the possibility that the dark matter particles are indeed unstable,
but decay leptonically with extremely long lifetimes. More precisely, the
interpretation of the leptonic cosmic-ray anomalies observed by PAMELA and
Fermi in terms of dark matter decay suggests a lifetime of the dark matter on
the order of $10^{26}$ seconds (see, e.g., ~\cite{Ibarra:2009dr}). This
lifetime exceeds the age of the Universe by nine orders of magnitude, thus
leaving the dark matter sufficiently stable on cosmological timescales.
Nevertheless, due to the large amounts of dark matter in the Universe, even
for such enormous lifetimes the resulting fluxes can be in the observable
range.  Indeed, strong constraints on decaying dark matter have been derived
recently from gamma-ray observations of galaxy clusters and nearby
galaxies~\cite{Dugger:2010ys}.

In this paper, we examine some of the effects of leptophilic models of
decaying dark matter which arise at next-to-leading order in perturbation
theory and show that they can have relevance to indirect dark matter searches.
We will not speculate on the precise nature of the particle physics that could
give rise to leptophilic dark matter decay. Instead, our approach will be to
examine simple models where we assume effective interactions that describe the
desired leptophilic coupling of dark matter particles to charged leptons. The
salient point for us here is that even if one assumes an exclusive coupling of
the dark matter to charged leptons at tree level, this behavior is only valid
at leading order, while at next-to-leading order other particles, including
photons and weak gauge bosons, will be produced. Indeed, these higher-order
corrections have been analyzed in the past for the case of annihilating dark
matter~\cite{Bergstrom:1989jr, Bringmann:2007nk, Dent:2008qy,
Kachelriess:2009zy, Bell:2010ei, Ciafaloni:2010ti}.  It is well known that the
higher-order corrections in the form of internal bremsstrahlung or from
final-state radiation of weak gauge bosons can even dominate under certain
conditions~\cite{Bringmann:2007nk,Bell:2010ei}.

The decay modes induced by higher-order corrections are usually suppressed by
powers of the couplings and possibly loop factors, as opposed to the
leading-order decay modes. This means that the resulting decay products will
be difficult to detect unless they possess some distinct features. Weak gauge
bosons can be produced, for instance, via final-state radiation off the
charged leptons~\cite{Berezinsky:2002hq}.  By their subsequent hadronization,
the massive gauge bosons will then generate hadronic particles, including
antiprotons~\cite{Ciafaloni:2010ti}.  Therefore, every leptophilic dark matter
model that aims to explain the leptonic cosmic-ray anomalies also serves as a
source of antiprotons.

In this work, however, we will focus on complementary constraints arising from
a different decay channel induced by higher-order effects. Namely, we will
study radiative two-body decays involving photons. These are particularly
interesting, since they give rise to monochromatic lines in the diffuse
gamma-ray spectrum or in extragalactic sources. Such lines are of utmost
importance because astrophysical processes generally generate continuous
gamma-ray spectra. Thus, the observation of a gamma-ray line would be a
compelling signature of an underlying particle physics process. In some cases,
a gamma-ray line can be produced already in tree level
decays~\cite{Ibarra:2007wg}. In the present work, we demonstrate that for
leptophilic models of dark matter, the ratio between leading-order and
next-to-leading-order decay modes can be large enough to produce a potentially
observable gamma-ray line signal.

This paper is organized as follows.  In section~\ref{sec:fermionicDM}, we
discuss the production of monochromatic photons from radiative two-body decays
induced at the one-loop level for fermionic dark matter particles in a simple
leptophilic toy model. In section~\ref{sec:scalarDM}, we examine the
corresponding case for a scalar dark matter particle. Next, in
section~\ref{sec:observational_constraints}, we discuss observational
constraints on gamma-ray lines in the GeV to TeV region and compare existing
bounds with the expected signal from dark matter decay. We also comment on
future observational prospects, in particular for the proposed Cherenkov
Telescope Array. Finally, we present our conclusions in
section~\ref{sec:Conclusions}.

\section{Radiative decay of fermionic dark matter}\label{sec:fermionicDM}

\begin{figure}
  \begin{center}
    \includegraphics{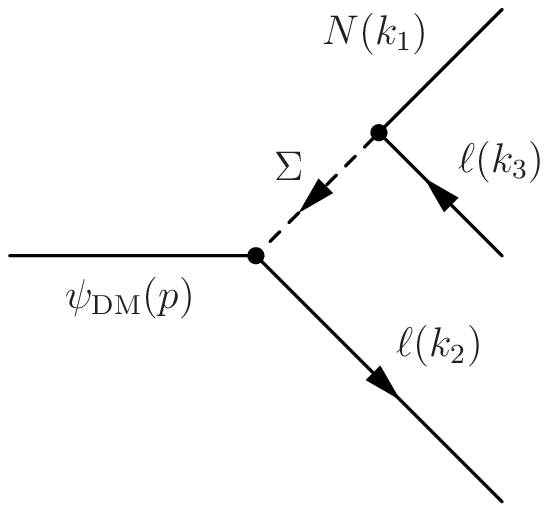}
    ~~~~~
    \includegraphics{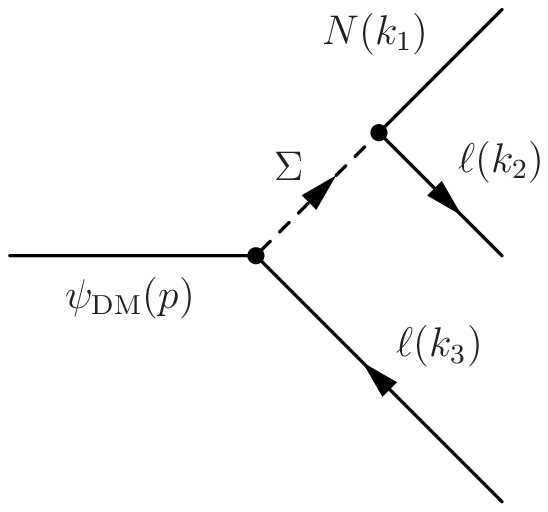}\\
    \caption{Tree-level diagrams contributing to the three-body decay
    $\psi_\text{DM} \rightarrow \ell^+ \ell^- N$ of fermionic dark matter,
    mediated by a heavy charged scalar $\Sigma$. Instead of the intermediate
    scalar $\Sigma$, the decay can also be mediated by a vector $V$.}
    \label{fermion_tree}
  \end{center}
\end{figure}

\begin{figure}
  \begin{center}
    \includegraphics{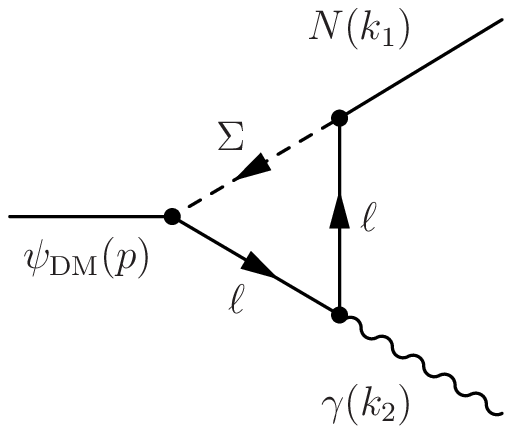}
    ~~~~~
    \includegraphics{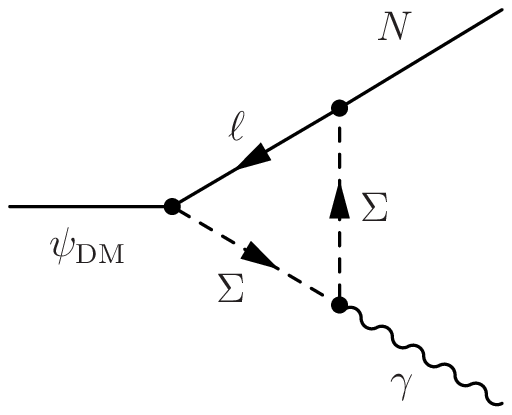}\\

    \

    \includegraphics{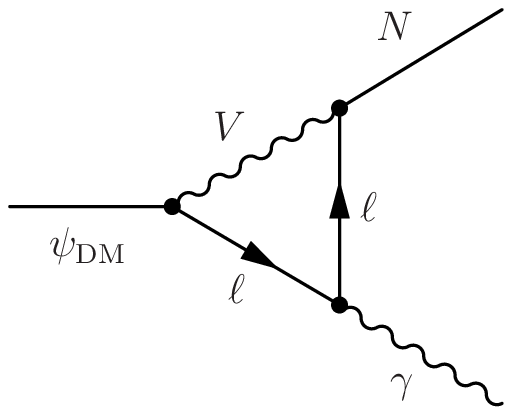}
    ~~~~~
    \includegraphics{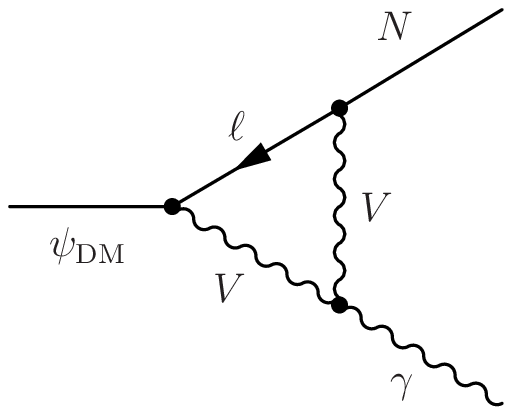}\\
    \caption{Diagrams contributing at one loop to the radiative two-body decay
    $\psi_\text{DM} \rightarrow \gamma N$, induced by a charged scalar
    $\Sigma$ (top row) and a vector particle $V$ (bottom row), respectively.
    There are two additional diagrams in each case which differ only by the
    direction of the charge flow.}
    \label{fermion_loop}
  \end{center}
\end{figure}

We first regard the case that the particles comprising the dark matter are
fermions which we denote by $\psi_\text{DM}$. We require that the dark matter
decays with a large branching fraction into pairs of charged leptons in order
to explain the excess of such leptons in high-energy cosmic rays, and we
assume that this is the only channel in which the dark matter decays at
leading order. If the dark matter carries spin 1/2,\footnote{For the case of 
spin 3/2, see refs.~\cite{Bajc:2010qj,Lola:2007rw} on radiative gravitino 
decay.} Lorentz invariance
requires the decay to be (at least) a three-body decay involving a third,
electrically neutral fermion $N$ for angular momentum conservation. Thus, the
decay $\psi_\text{DM} \rightarrow \ell^+ \ell^- N$ is the simplest one allowed
by gauge and Lorentz invariance. Here, $N$ could be a neutrino, a neutralino
or a gravitino, for instance. The decays may be mediated by a virtual charged
scalar particle $\Sigma$ or by a charged vector boson $V$, with masses
$m_\Sigma$ and $m_V$, respectively, which are assumed to be larger than the
mass of the dark matter particle. We regard the two cases separately.

In the case of an intermediate scalar, the effective Lagrangian that we use is
given as the sum of a term coupling the dark matter, which we take to be a
metastable Majorana fermion, to a charged lepton and a $\Sigma$ particle, as
well as a term coupling the neutral fermion to the $\Sigma$ and a lepton
field. We decompose the couplings into left- and right-handed components to
allow for chiral couplings. Then the Lagrangian has the form
\begin{align}
  \mathcal{L}_\text{eff}^\Sigma = -\bar{\psi}_\text{DM} \left[\lambda_{\ell
  \psi}^L P_L + \lambda_{\ell \psi}^R P_R\right] \ell \, \Sigma^\dagger
  -\bar{N} \left[\lambda_{\ell N}^L P_L + \lambda_{\ell N}^R P_R\right] \ell
  \, \Sigma^\dagger + \text{h.c.}\;,
\end{align}
where $P_L = (1 - \gamma^5)/2$ and $P_R = (1 + \gamma^5)/2$ are the left- and
right-handed chirality projectors, respectively. The $\lambda$-couplings can
in general be complex. To obtain the required cosmological lifetime for the
dark matter, the couplings have to be super-weak or the mass $m_\Sigma$ of the
mediator has to be super-heavy. The operators of the effective Lagrangian
induce three-body decays of the dark matter into a pair of charged leptons and
a neutral fermion at tree level, $\psi_\text{DM} \rightarrow \ell^+ \ell^- N$.
The corresponding diagrams are shown in fig.~\ref{fermion_tree}.

In the case of a vector interaction, on the other hand, we assume an effective
Lagrangian of the form
\begin{equation}
  \mathcal{L}_\text{eff}^V = -\bar{\psi}_\text{DM} \gamma^\mu
  \left[\lambda_{\ell \psi}^L P_L + \lambda_{\ell \psi}^R P_R\right] \ell \,
  V_\mu^\dagger - \bar{N} \gamma^\mu \left[\lambda_{\ell N}^L P_L +
  \lambda_{\ell N}^R P_R\right] \ell \, V_\mu^\dagger + \text{h.c.}\;.
\end{equation}
The case of mediation by a vector boson is more involved than the previous
case of mediation by a scalar. In choosing a Lagrangian of this form, we
assume that the essence of the gauge interaction giving rise to this
Lagrangian is captured by the effective charged-vector interaction. In
general, one expects neutral currents in association with the charged
currents, which introduces a high degree of model dependence. For simplicity,
we assume here that the decay is dominated by the charged-current interaction.

\subsection{Decay widths}
In the following we examine decay modes of the dark matter at the tree- and
one-loop level and summarize the relevant decay widths.\footnote{We have
cross-checked the matrix elements for the three-body decays and the decay
rates in the following sections by comparing them to the results from
FeynArts~\cite{Hahn:2000kx} and FormCalc~\cite{Hahn:1998yk}.}

\subsubsection{Tree-level decay: \boldmath$\psi_\text{DM} \rightarrow \ell^+
\ell^- N$}
The leading-order decay induced by the effective Lagrangian is the three-body
decay $\psi_\text{DM} \rightarrow \ell^+ \ell^- N$. If the dark matter decays
in this way, it constitutes a possible explanation for the observed cosmic-ray
anomalies under certain conditions~\cite{Ibarra:2009dr}. We present the
relevant expressions for the decay widths in the following.

\paragraph{Mediation by a scalar.}
In the plausible limit $m_\ell \ll m_{\psi_\text{DM}} \ll m_\Sigma$, the
partial decay width for the decay $\psi_\text{DM} \rightarrow \ell^+ \ell^- N$
is given by (see app.~\ref{app:fermion} and ref.~\cite{Bartl:1986hp})
\begin{equation}\label{eqn:fermion_3body_width}
  \Gamma(\psi_\text{DM} \rightarrow \ell^+ \ell^- N) = \frac{1}{64 (2 \pi)^3}
  \frac{m_{\psi_\text{DM}}^5}{6 m_\Sigma^4} \left\{C_1^\Sigma
  F_1(m_N^2/m_{\psi_\text{DM}}^2) + C_2^\Sigma
  F_2(m_N^2/m_{\psi_\text{DM}}^2)\right\}.
\end{equation}
The constants $C_1^\Sigma$, $C_2^\Sigma$ are determined by the couplings as
\begin{align}
  C_1^\Sigma &\equiv \left(|\lambda_{\ell \psi}^L|^2 + |\lambda_{\ell
  \psi}^R|^2\right) \left(|\lambda_{\ell N}^L|^2 + |\lambda_{\ell
  N}^R|^2\right) - \eta\,\text{Re} \left(\lambda_{\ell
  \psi}^L \lambda_{\ell N}^{L*} \lambda_{\ell \psi}^R \lambda_{\ell
  N}^{R*}\right),\\
  C_2^\Sigma &\equiv 2 \eta\,
  \text{Re}\left[\left(\lambda_{\ell \psi}^L \lambda_{\ell N}^{L*}\right)^2 +
  \left(\lambda_{\ell \psi}^R \lambda_{\ell N}^{R*}\right)^2\right].
\end{align}
Here, $\eta \equiv \eta_{\psi_\text{DM}} \eta_N = \pm 1$ depending on the
\textsl{CP} eigenvalues of $\psi_\text{DM}$ and $N$. The kinematical
functions, on the other hand, are given by
\begin{align}
  F_1(x) &\equiv (1 - x^2)(1 + x^2 - 8x) - 12x^2 \ln(x), \label{eqn:F1}\\
  F_2(x) &\equiv \sqrt{x} [(1 - x) (1 + 10x + x^2) + 6x (1 + x) \ln(x)].
  \label{eqn:F2}
\end{align}
In the hierarchical limit $m_N/m_{\psi_\text{DM}} \rightarrow 0$, the
kinematical functions satisfy
\begin{equation}
  F_1(x) \simeq 1, ~~~ F_2(x) \simeq \sqrt{x} ~~~ \text{for} ~~~ x \rightarrow
  0,
\end{equation}
whereas in the degenerate limit $m_N/m_{\psi_\text{DM}} \rightarrow 1$, one
gets
\begin{equation}
  F_1(x) \simeq \frac{2}{5}(1 - x)^5, ~~~ F_2(x) \simeq \frac{1}{10}(1 - x)^5
  ~~~ \text{for} ~~~ x \rightarrow 1.
\end{equation}
In the limit $m_N \ll m_{\psi_\text{DM}}$ the decay rate
(\ref{eqn:fermion_3body_width}) corresponds to a lifetime
\begin{equation}\label{LifetimeThreeBodyScalar}
  \tau_{\psi_\text{DM} \rightarrow \ell^+ \ell^- N} \simeq 6 \times
  10^{26}\,\text{s}\left(\frac{0.1}{C_1^\Sigma}\right)
  \left(\frac{1~\text{TeV}}{m_{\psi_\text{DM}}}\right)^5
  \left(\frac{m_\Sigma}{10^{15}~\text{GeV}}\right)^4.
\end{equation}
In the case where $m_{\psi_\text{DM}}$ and $m_N$ are quasi-degenerate, the
decay rate scales approximately like $(m_{\psi_\text{DM}} - m_N)^5$.

\paragraph{Mediation by a vector.}
For the vector-mediated decay we find for the three-body decay rate in the
limit $m_\ell \ll m_{\psi_\text{DM}} \ll m_V$
\begin{equation}
  \Gamma(\psi_\text{DM} \rightarrow \ell^+ \ell^- N) = \frac{1}{64 (2 \pi)^3}
  \frac{4 m_{\psi_\text{DM}}^5}{6 m_V^4} \left\{C_1^V
  F_1(m_N^2/m_{\psi_\text{DM}}^2) + C_2^V
  F_2(m_N^2/m_{\psi_\text{DM}}^2)\right\}\;,
\end{equation}
where the functions $F_1$ and $F_2$ are the same as in eqs.~(\ref{eqn:F1}),
(\ref{eqn:F2}) and
\begin{align}
  C_1^V &\equiv \left(|\lambda_{\ell \psi}^L|^2 + |\lambda_{\ell
  \psi}^R|^2\right) \left(|\lambda_{\ell N}^L|^2 + |\lambda_{\ell
  N}^R|^2\right) + 2 \eta\,\text{Re}
  \left(\lambda_{\ell \psi}^L \lambda_{\ell N}^{L*} \lambda_{\ell \psi}^R
  \lambda_{\ell N}^{R*}\right),\\
  C_2^V &\equiv 2 \eta\,
  \text{Re}\left[\left(\lambda_{\ell \psi}^L \lambda_{\ell N}^{L*}\right)^2 +
  \left(\lambda_{\ell \psi}^R \lambda_{\ell N}^{R*}\right)^2\right] =
  C_2^\Sigma.
\end{align}
For $m_N \ll m_{\psi_\text{DM}}$ and an analogous choice of parameters, the
lifetime is smaller by a factor of four compared to
eq.~(\ref{LifetimeThreeBodyScalar}),
\begin{equation}\label{LifetimeThreeBodyVector}
  \tau_{\psi_\text{DM} \rightarrow \ell^+ \ell^- N} \simeq 1.5 \times
  10^{26}\,\text{s}\left(\frac{0.1}{C_1^V}\right)
  \left(\frac{1~\text{TeV}}{m_{\psi_\text{DM}}}\right)^5
  \left(\frac{m_V}{10^{15}~\text{GeV}}\right)^4.
\end{equation}

\subsubsection{One-loop decay: \boldmath$\psi_\text{DM} \rightarrow \gamma N$} 
By combining the external charged lepton lines from the tree-level diagrams
into a loop, we obtain diagrams contributing to the two-body decay
$\psi_\text{DM} \rightarrow \gamma N$ (see fig.~\ref{fermion_loop}). This
decay mode will be suppressed with respect to the tree-level three-body decay
by a loop factor and an additional power of the electromagnetic coupling. This
is partially compensated by phase-space factors, however.  More importantly,
the two-body decay gives rise to monochromatic photons at an energy
\begin{equation}\label{eqn:E_gamma}
  E_\gamma = \frac{m_{\psi_\text{DM}}}{2} \left(1 -
  \frac{m_N^2}{m_{\psi_\text{DM}}^2}\right)\;,
\end{equation}
which can result in a distinct observational signature at gamma-ray
telescopes, as will be discussed in some detail in
section~\ref{sec:observational_constraints}. Interestingly, the experimental
constraints on the parameters of decaying dark matter stemming from the
non-observation of energetic gamma-ray lines could, despite the
loop-suppression, be more stringent than the ones stemming from measurements
of cosmic-ray electrons and positrons. 

Based on gauge invariance, and irrespective of whether the decay is mediated
by a scalar or a vector particle, the matrix element for the sum of all
diagrams contributing to the radiative two-body decay can be written in the
following form, introducing an effective coupling $g_{N \gamma
\psi}$~\cite{Haber:1988px},
\begin{align}
  \mathcal{M} &= \frac{i g_{N \gamma \psi}}{m_{\psi_\text{DM}}}
  \bar{u}(k_1)(P_R - \eta_N \eta_{\psi_\text{DM}} P_L) \sigma^{\mu \nu} k_{2
  \mu} \epsilon_\nu^* u(p)\nonumber \\ &= -\frac{g_{N \gamma
  \psi}}{m_{\psi_\text{DM}}} \bar{u}(k_1)(P_R - \eta_N \eta_{\psi_\text{DM}}
  P_L) \slashed{k}_2 \slashed{\epsilon}^* u(p) \label{eqn:ME_two_body}\;,
\end{align}
where $\sigma^{\mu \nu} = i [\gamma^\mu,\gamma^\nu]/2$ and
$\eta_{\psi_\text{DM}}$, $\eta_N$ are the \textsl{CP} eigenvalues of
$\psi_\text{DM}$ and $N$, respectively.  The partial decay width for
$\psi_\text{DM} \rightarrow \gamma N$ can then be easily calculated to be
\begin{equation}\label{eqn:fermion_2body_width}
  \Gamma(\psi_\text{DM} \rightarrow \gamma N) = \frac{g_{N \gamma \psi}^2}{8
  \pi} m_{\psi_\text{DM}} \left(1 -
  \frac{m_N^2}{m_{\psi_\text{DM}}^2}\right)^3.
\end{equation}
The effective coupling $g_{N \gamma \psi}$ encodes all the information about
the interaction between dark matter and the decay products. We give explicit
expressions for this coupling in the following.

\paragraph{Mediation by a scalar.}
We first examine the case of mediation by a charged scalar particle $\Sigma$
(top row of fig.~\ref{fermion_loop}). Assuming that \textsl{CP} is conserved
in the interactions of $\psi_\text{DM}$ and $N$, i.e., when the
$\lambda$-couplings are assumed to be real, the explicit form of the effective
coupling $g_{N \gamma \psi}^\Sigma$ can be expressed as follows,
\begin{align}
  g_{N \gamma \psi}^\Sigma = &-\frac{e \, \eta_N m_{\psi_\text{DM}}}{16 \pi^2}
  \sum_{\ell,\Sigma} Q_\ell C_\ell \Big\{m_f (\eta_{\psi_\text{DM}} \lambda_{\ell N}^L
  \lambda_{\ell \psi}^R - \eta_N \lambda_{\ell N}^R \lambda_{\ell \psi}^L) I
  \nonumber\\ &+ (\lambda_{\ell N}^L \lambda_{\ell \psi}^L -
  \eta \, \lambda_{\ell N}^R \lambda_{\ell \psi}^R)
  [\eta_{\psi_\text{DM}} m_{\psi_\text{DM}} (I^2 - K) - \eta_N m_N K]\Big\}\;,
\end{align}
where the loop integrals $I$, $I^2$ and $K$ are defined in
app.~\ref{app:fermion}.\footnote{Note that the superscript `2' in the integral
$I^2$ is an index, not a square.} The sum runs over all lepton flavors
$\ell\in\{e,\mu,\tau\}$, for which $Q_\ell = C_\ell = 1$. If multiple mediator
particles (like left- and right-handed sleptons, $\Sigma =
\widetilde{\ell}_{L},\widetilde{\ell}_{R}$) are present, one also has to sum
over their contributions. In principle, there can also be contributions from
quarks in the loop. Then, the sum runs over quarks and leptons with electric
charge $Q_q$ and color charge $C_q = 3$.  However, tree-level decays into
quarks can potentially lead to an overproduction of antiprotons if the
relative size of the effective coupling to quarks compared to the coupling to
leptons is too large. The requirement of avoiding antiproton overproduction
then leads to the assumption of a leptophilic structure. For this reason, we
assume throughout this work that the dark matter decays only into leptons at
tree level.

If the mass of the intermediate particle $\Sigma$ is much larger than the
other masses, the loop integrals take on a very simple form. The effective
coupling is then given approximately by
\begin{equation}
  g_{N \gamma \psi}^\Sigma \simeq \frac{e \, \eta}{64\pi^2} m_{\psi_\text{DM}}^2
  \left(1 - \frac{\eta \, m_N}{m_{\psi_\text{DM}}}\right) \sum_{\ell,\Sigma}
  \frac{Q_\ell C_\ell}{m_\Sigma^2} \left\{\left(\lambda_{\ell N}^L
  \lambda_{\ell \psi}^L - \eta \, \lambda_{\ell N}^R \lambda_{\ell
  \psi}^R\right)\right\}\;.
\end{equation}

For the concrete case where there is only one mediator $\Sigma$, which couples
exclusively to leptons, the decay rate reads
\begin{align}
  \Gamma(\psi_\text{DM} \rightarrow \gamma N) = {}& \frac{e^2}{8\pi
  \left(64\pi^2\right)^2} \frac{m_{\psi_\text{DM}}^5}{m_\Sigma^4} \left(1 -
  \frac{m_N^2}{m_{\psi_\text{DM}}^2}\right)^3 \left(1 - \frac{\eta \,
  m_N}{m_{\psi_\text{DM}}}\right)^2 \nonumber\\ & \times \left[\sum_\ell
  \left(\lambda_{\ell N}^L \lambda_{\ell \psi}^L - \eta \, \lambda_{\ell N}^R
  \lambda_{\ell \psi}^R\right)\right]^2\;.
\end{align}
For $m_N\ll m_{\psi_\text{DM}}$, this decay width corresponds to a partial
lifetime
\begin{equation}
  \tau_{\psi_\text{DM} \rightarrow \gamma N} \simeq 7 \times 10^{29}
  ~\text{s}~ \frac{0.1}{\left[\left(\lambda_{\ell N}^L \lambda_{\ell \psi}^L
  - \eta \, \lambda_{\ell N}^R \lambda_{\ell \psi}^R\right)\right]^2}
  \left(\frac{1~\text{TeV}}{m_{\psi_\text{DM}}}\right)^5
  \left(\frac{m_\Sigma}{10^{15}~\text{GeV}}\right)^4\;.
\end{equation}

\paragraph{Mediation by a vector.}
In the case of mediation by a charged vector boson (bottom row of
fig.~\ref{fermion_loop}) we obtain the following expression for the effective
coupling~\cite{Haber:1988px},
\begin{align}
  g_{N \gamma \psi}^V = {}& \frac{e \, \eta_N m_{\psi_\text{DM}}}{8 \pi^2}
  \sum_{\ell} \Big\{(\eta_{\psi_\text{DM}} \eta_N \lambda_{\ell N}^L \lambda_{\ell
  \psi}^L - \lambda_{\ell N}^R \lambda_{\ell \psi}^R)
  \big[\eta_{\psi_\text{DM}} m_{\psi_\text{DM}} (I^2 - J - K) \nonumber\\ & +
  \eta_N m_N (J - K)\big] + 2 m_\ell (\eta_{\psi_\text{DM}} \lambda_{\ell N}^L
  \lambda_{\ell \psi}^R - \eta_N \lambda_{\ell N}^R \lambda_{\ell \psi}^L)
  J\Big\}\;,
\end{align}
where we encounter an additional loop integral $J$, which is defined in
app.~\ref{app:fermion}.  Again, it is possible to include quarks by the
replacement $\sum_\ell \rightarrow \sum_f Q_f C_f$. However, as discussed
above, this would not correspond to a leptophilic model.  In the limit $m_\ell
\rightarrow 0$, $m_{\psi_\text{DM}} \ll m_V$, the above expression simplifies
to
\begin{align}
  g_{N \gamma \psi}^V \simeq \frac{3e \, \eta}{32\pi^2}
  \frac{m_{\psi_\text{DM}}^2}{m_V^2} \left(1 - \frac{\eta \,
  m_N}{m_{\psi_\text{DM}}}\right) \sum_\ell \left(\lambda_{\ell N}^L
  \lambda_{\ell \psi}^L - \eta \, \lambda_{\ell N}^R \lambda_{\ell
  \psi}^R\right)\;.
\end{align}
Thus, in the limit $m_\ell \ll m_N$ and $m_{\psi_\text{DM}} \ll m_V$ we obtain
for the decay width
\begin{align}
  \Gamma(\psi_\text{DM} \rightarrow \gamma N) &= \frac{9e^2}{8\pi
  \left(32\pi^2\right)^2} \frac{m_{\psi_\text{DM}}^5}{m_V^4} \left(1 -
  \frac{m_N^2}{m_{\psi_\text{DM}}^2}\right)^3 \left(1 - \frac{\eta \,
  m_N}{m_{\psi_\text{DM}}}\right)^2 \nonumber\\ & ~~~ \times \left[\sum_\ell
  \left(\lambda_{\ell N}^L \lambda_{\ell \psi}^L - \eta \, \lambda_{\ell N}^R
  \lambda_{\ell \psi}^R\right)\right]^2\;.
\end{align}
For $m_N\ll m_{\psi_\text{DM}}$, this yields a partial lifetime
\begin{align}
  \tau_{\psi_\text{DM} \rightarrow \gamma N} \simeq 2 \times 10^{28}
  ~\text{s}~ \frac{0.1}{\left[\sum_\ell \left(\eta \, \lambda_{\ell N}^L
  \lambda_{\ell \psi}^L - \lambda_{\ell N}^R \lambda_{\ell
  \psi}^R\right)\right]^2}
  \left(\frac{1~\text{TeV}}{m_{\psi_\text{DM}}}\right)^5
  \left(\frac{m_V}{10^{15}~\text{GeV}}\right)^4\;.
\end{align}

\subsection{Intermediate scalar: intensity of the gamma-ray line}
The detectability of a loop-induced gamma-ray line will depend crucially on
the ratio between the three-body decays at tree level and the two-body decays
at the loop level. We examine the general expressions first and then evaluate
them for some specific examples.

\subsubsection{General expressions}\label{sec:scalar_intensity}
In the intermediate scalar case, the ratio between two- and three-body decay
widths reads, neglecting the charged lepton masses,
\begin{align}
  \label{ratio-scalar} \frac{\Gamma(\psi_\text{DM} \rightarrow \gamma N
  )}{\sum_\ell \Gamma(\psi_\text{DM} \rightarrow \ell^+ \ell^- N)}
  \simeq \frac{3 \alpha_\text{em}}{8 \pi} \frac{\left[\sum_\ell
  \left(\lambda_{\ell N}^L \lambda_{\ell \psi}^L - \eta \, \lambda_{\ell N}^R
  \lambda_{\ell \psi}^R\right)\right]^2 (1 - x)^3 (1 - \eta
  \sqrt{x})^2}{\sum_\ell C_1^\Sigma F_1(x) + C_2^\Sigma F_2(x)},
\end{align}
where $x \equiv m_N^2/m_{\psi_\text{DM}}^2$ and the kinematical functions
$F_1$ and $F_2$ were defined in eqs.~(\ref{eqn:F1}), (\ref{eqn:F2}). This
general expression can be used to study the intensity of the one-loop induced
gamma-ray line in different scenarios. The numerical value of the prefactor is
$3\alpha_\text{em}/(8 \pi) \simeq 1/1148$.

In general, the fraction depends on the chiral and flavor structure of the
couplings, the mass ratio $m_\nn/m_\fdm$ of the decay product and the dark
matter particle, and the relative \textsl{CP} parities $\eta=\pm 1$ of $\nn$
and $\fdm$. For many practical purposes, it turns out that the dependence on
the couplings $\lambda_{\ell\nn/\psi}$ and on kinematics, i.e. on $x =
m_N^2/m_{\psi_\text{DM}}^2$, can be factored according to
\begin{align}\label{ratio-scalar-RS}
  \frac{\Gamma(\psi_\text{DM} \rightarrow \gamma N)}{\sum_\ell
  \Gamma(\psi_\text{DM} \rightarrow \ell^+ \ell^- N)} \simeq \frac{3
  \alpha_\text{em}}{8 \pi} \times R_\eta^\Sigma(\lambda_{\ell
  N}^L,\lambda_{\ell \psi}^L,\lambda_{\ell N}^R,\lambda_{\ell \psi}^R) \times
  S_\eta(m_N/m_{\psi_\text{DM}}) \,.
\end{align}
In this parametrization $R_\eta^\Sigma$ captures the model-dependence, whereas
$S_\eta$ is determined entirely by kinematics.

It is interesting to consider the two limiting cases of hierachical masses,
$m_\nn/m_\fdm\to 0$, and degenerate masses, $m_\nn/m_\fdm\to 1$.  In the
hierarchical limit, and assuming for simplicity real couplings, one explicitly
obtains $S_\eta^{\rm hier}= 1$ and
\begin{align}
  R^{\Sigma, {\rm hier}}_\eta(\lambda_{\ell N}^L,\lambda_{\ell
  \psi}^L,\lambda_{\ell N}^R,\lambda_{\ell \psi}^R) =  \frac{\left[\sum_\ell
  \left(\lambda_{\ell N}^L \lambda_{\ell \psi}^L - \eta \, \lambda_{\ell N}^R
  \lambda_{\ell \psi}^R\right)\right]^2}{\sum_\ell \left[\big(\lambda_{\ell
  \psi}^{L2} +\lambda_{\ell \psi}^{R2} \big) \big(\lambda_{\ell N}^{L2}
  +\lambda_{\ell N}^{R2}\big) - \eta \, \lambda_{\ell \psi}^L \lambda_{\ell N}^L
  \lambda_{\ell \psi}^R \lambda_{\ell N}^R\right]}\;.
\end{align}
For generic couplings, $R^{\Sigma, {\rm hier}}_\eta$ is roughly of order one,
unless for some special cases where cancellations or chirality suppressions
occur. It follows that in the hierarchical limit the two-body decays into
$\gamma N$ are typically suppressed roughly by a factor $10^{-3}$ compared to
the tree-level decays into $\ell^+\ell^-\nn$. In the next subsection we will
examine the model-dependent factor $R_\eta^\Sigma$ for some specific cases.

\begin{figure}
  \begin{center}
    \includegraphics[width=12cm]{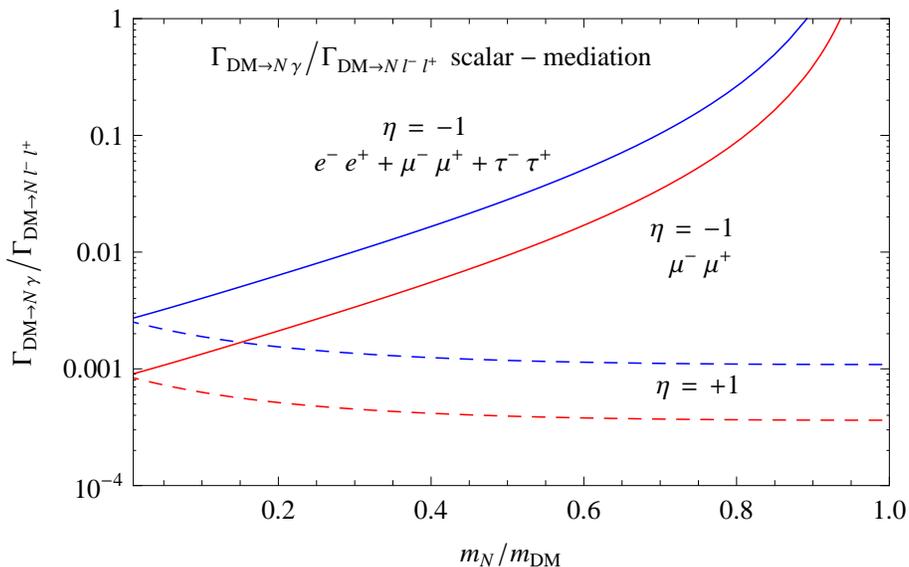}
  \end{center}
  \caption{Ratio of the decay rates $\Gamma(\psi_\text{DM} \rightarrow \gamma
  N)/\sum_\ell \Gamma(\psi_\text{DM} \rightarrow \ell^+ \ell^- N)$ when the
  decay is mediated by a scalar. The four cases correspond to single-flavor
  decay (red) and democratic decay into all flavors (blue), as well as
  $\fdm/\nn$ having the same \textsl{CP} parity (dashed, $\eta=+1$ ) or
  opposite \textsl{CP} parity (solid, $\eta=-1$ ).  See eq.
  (\ref{ratioScalar}).}\label{fig:ratio}
\end{figure}

On the other hand, in the degenerate limit $m_\nn/m_\fdm\to 1$, and again
assuming real couplings, one finds
\begin{align}
  R^{\Sigma, {\rm deg}}_\eta(\lambda_{\ell N}^L,&\lambda_{\ell
  \psi}^L,\lambda_{\ell N}^R,\lambda_{\ell \psi}^R) =
  \\\nonumber
  &\frac{\left[\sum_\ell
  \left(\lambda_{\ell N}^L \lambda_{\ell \psi}^L - \eta \, \lambda_{\ell N}^R
  \lambda_{\ell \psi}^R\right)\right]^2 }{\sum_\ell
  \left[\frac{2}{2+\eta}\big(\lambda_{\ell \psi}^{L2} +\lambda_{\ell
  \psi}^{R2} \big) \big(\lambda_{\ell N}^{L2} +\lambda_{\ell N}^{R2}\big) +
  \frac{\eta}{2+\eta}\,\left(\lambda_{\ell \psi}^L \lambda_{\ell N}^L
  -\lambda_{\ell \psi}^R \lambda_{\ell N}^R\right)^2\right]}\;,
\end{align}
which is also roughly of order one for generic couplings, $R^{\Sigma, {\rm
deg}}_\eta \sim \mathcal{O}(1)$, and
\begin{align}\label{S}
  S^{\rm deg}_\eta(m_N/m_{\psi_\text{DM}})\simeq \left\{\begin{array}{ccl}
  5/12 & {\rm for} & \eta=+1 \\ 20/(1-m_N^2/m_{\psi_\text{DM}}^2)^2 & {\rm
  for} & \eta=-1 \;.  \end{array}\right.
\end{align}
Thus, for the case $\eta=+1$, i.e. when $\fdm$ and $\nn$ have the same
\textsl{CP} parities, we again find a typical suppression factor of the order
of $10^{-3}$ for the two-body relative to the tree-level decay rate, as in the
hierarchical case. Interestingly, however, when $\fdm$ and $\nn$ have opposite
\textsl{CP} parities, $\eta=-1$, the two-body rate can be enhanced
significantly even for a relatively mild degeneracy, as is shown in
fig.~\ref{fig:ratio}.  This enhancement is due to the fact that the decay rate
$\Gamma(\fdm\to \gamma N)$ is proportional to $(m_\nn-m_\fdm)^3$, whereas the
decay into leptons is suppressed like $(m_\nn-m_\fdm)^5$~\cite{Haber:1988px}.
Most interestingly, due to this enhancement the decay rate into $\gamma N$ can
be rather large in some cases, yielding potentially very intense gamma-ray
lines.

As a side remark, we note that in addition to the decay channel $\fdm\to\gamma
N$ into photons, there can exist a decay mode $\fdm\to Z^0 N$ into $Z$-bosons,
which can naively be expected to be of similar size.  Thus, for situations
where the gamma-ray line signal is strongly enhanced, an equally enhanced
decay into $Z$-bosons can yield additional constraints from the antiproton
flux produced by the subsequent fragmentation of the $Z$-bosons. We leave a
more detailed discussion for the future \cite{future}.

For concreteness, we consider the case of purely chiral, say left-handed,
couplings, and that only one mediator species is present. Then the fraction of
decay rates is given by
\begin{align}
  \label{ratioScalar} \frac{\Gamma(\psi_\text{DM} \rightarrow \gamma N
  )}{\sum_\ell \Gamma(\psi_\text{DM} \rightarrow \ell^+ \ell^- N)}
  \simeq \frac{3 \alpha_\text{em}}{8 \pi} R^{\rm chir} \,
  \left\{\begin{array}{ll} 1 & \mbox{for } m_\nn \to 0, ~ \eta = \pm 1\\
    \frac{5}{12} & \mbox{for } m_\nn \to m_\fdm, ~ \eta = +1\\
    \frac{20}{(1-m_\nn^2/m_\fdm^2)^2} & \mbox{for } m_\nn \to m_\fdm, ~ \eta =
    -1 \end{array}\right. \;.
\end{align}
In this case the model-dependent factor for the hierarchical and degenerate
regimes coincides, $R^{\Sigma, {\rm hier}}=R^{\Sigma, {\rm deg}}\equiv R^{\rm
chir}$, and is furthermore independent of $\eta$. Explicitly, one has
\begin{equation}\label{Rchir}
  R^{\rm chir} = \frac{\left[\sum_\ell \lambda_{\ell N}^L \lambda_{\ell
  \psi}^L \right]^2 }{\sum_\ell \left( \lambda_{\ell N}^L \lambda_{\ell
  \psi}^L \right)^2 } \;.
\end{equation}
The result for purely right-handed couplings is analogous.  For a generic
choice of couplings, one expects $R^{\rm chir} \sim \mathcal{O}(1)$. Note that
$R^{\rm chir} \leq N_\ell$, where $N_\ell$ is the number of flavors
participating in the decay.

For example, consider two particular cases for the flavor composition of the
lepton pairs produced in the decay:
\begin{itemize}
  \item[(A)] Decay into a single lepton flavor: $\mu^+\mu^-$,
  \item[(B)] Flavor-democratic decay into $e^+e^-$, $\mu^+\mu^-$,
    $\tau^+\tau^-$.
\end{itemize}
Then one has $R^{\rm chir}=1$ in case (A) and $R^{\rm chir}=3$ in case (B).

The dependence of the ratio of decay rates on the mass ratio $m_\nn/m_\fdm$ is
shown in fig.~\ref{fig:ratio} for the two cases (A) and (B), and for $\eta=\pm
1$.  This dependence is in fact a rather generic feature, which is independent
of the details of the couplings. We emphasize again that, in the case when
$\fdm$ and $\nn$ have opposite \textsl{CP} parities, $\eta=-1$, even a rather
mild degeneracy between $m_\nn$ and $m_\fdm$ can lead to a considerable
enhancement of the gamma-ray line signal relative to the electron/positron
flux.

\subsubsection{Examples}\label{sec:Examples}
Now, we will discuss the branching ratio into gamma-ray lines for several
specific scenarios. Namely, we consider the case where $N$ corresponds to
left-handed neutrinos $\nu_L$, as well as the scenario of kinetically mixed
hidden $U(1)$ gauginos, where $\nn$ corresponds to a neutralino.\\

\paragraph{Decay into left-handed neutrinos.}\

As a basic example for the scalar-mediated decay described in the previous
subsection, we consider the case where the neutral fermion is a left-handed
neutrino, $N \equiv \nu_L$.  Then one can set $\lambda_{\ell \nu}^L = 0$ and
$m_\nu = 0$. 

From eq.~\eqref{ratio-scalar} it directly follows, that in the limit $m_\ell
\ll m_{\psi_\text{DM}} \ll m_\Sigma$ the ratio reads
\begin{equation}
  \frac{\Gamma(\psi_\text{DM} \rightarrow \gamma \nu)}{\sum_\ell
  \Gamma(\psi_\text{DM} \rightarrow \ell^+ \ell^- \nu)} \simeq \frac{3
  \alpha_\text{em}}{8 \pi} \frac{\left[\sum_\ell \lambda_{\ell \nu}^R
  \lambda_{\ell \psi}^R\right]^2}{\sum_\ell \left(\left|\lambda_{\ell
  \psi}^L\right|^2 + \left|\lambda_{\ell \psi}^R\right|^2\right)
  \left|\lambda_{\ell \nu}^R\right|^2}.
\end{equation}
As long as only one virtual scalar particle is relevant for the three-body
decay, the last factor in this expression is bounded from above by the number
of lepton flavors $N_\ell$ that contribute to the decay. For $N_\ell = 3$,
this corresponds to a branching ratio into monochromatic photons smaller than
$3\times 10^{-3}$.

\paragraph{Hidden-gaugino dark matter.}\

In supersymmetric scenarios with an extra $U(1)_X$ gauge group in the hidden
sector, which kinetically mixes with the Standard Model $U(1)_Y$, the
particles $\psi_\text{DM}$ and $N$ could be associated with the hidden gaugino
and the lightest MSSM neutralino, respectively. If the kinetic mixing
parameter $\theta$ is extremely small, the hidden gaugino could constitute
decaying dark matter~\cite{Ibarra:2008kn}. For a bino-like lightest neutralino
and $\Sigma \equiv \widetilde{\ell}_L$ being a left-handed slepton, the
couplings are approximately given by
\begin{align}
  \lambda_{\ell \psi}^L &\simeq \frac{g'}{\sqrt{2}} Y_\ell^L \theta,\\
  \lambda_{\ell N}^L &\simeq \frac{g'}{\sqrt{2}} Y_\ell^L,
\end{align}
where $\theta \sim 10^{-24}$ is the mixing angle of hidden gaugino and bino,
fixed by the requirement of a lifetime of the order of $10^{26}$ s, and
$Y_\ell^L = +1$.

For a hidden gaugino that decays into a bino-like neutralino, the decay rates
are given in app.~A of ref.~\cite{Ibarra:2008kn}. One can also obtain these
rates using the expressions derived above. In particular, we have to sum over
two `mediators' $\med$ for each flavor:
\begin{itemize}
  \item $\med = \widetilde\ell_L$: \ $\lamFL =
    \frac{g'}{\sqrt{2}}Y_\ell^L\theta$, \ $\lamNNL =
    \frac{g'}{\sqrt{2}}Y_\ell^L$, \ $\lamFR \simeq 0$, \ $\lamNNR \simeq 0$
  \item $\med = \widetilde\ell_R$: \ $\lamFR =
    \frac{g'}{\sqrt{2}}Y_\ell^R\theta$, \ $\lamNNR =
    \frac{g'}{\sqrt{2}}Y_\ell^R$, \ $\lamFL \simeq 0$, \ $\lamNNL \simeq 0$\;.
\end{itemize}
In addition, there are corresponding contributions from (s)quarks. We assume
$m_{\widetilde f} \gg m_{\fdm}$, and neglect the mixing of $\widetilde
f_{L,R}$ for simplicity.  Note that there are additional contributions from
chargino loops~\cite{Haber:1988px} that are suppressed by the fourth power of
the inverse chargino mass. We assume that the squarks are much heavier than
the sleptons, and we furthermore assume that all slepton masses are
degenerate.  Finally, if we take the limit $m_\ell \rightarrow 0$, we get
\begin{eqnarray}
  \Gamma(\fdm\rightarrow \gamma\nn) & \simeq &
  \frac{e^2{g'}^4\theta^2}{8\pi\left(32\pi^2\right)^2}
  \frac{m_{\fdm}^5}{16m_{\widetilde \ell}^4} \left(1 -
  \frac{m_{\nn}^2}{m_{\fdm}^2}\right)^3 \left(1 -
  \eta \, \frac{m_{\nn}}{m_{\fdm}}\right)^2 \nonumber\\ &   & \times \left[ 3
  \left( 1 - 4\eta\right) \right]^2 \;.
\end{eqnarray}
If the bino and the hidden gaugino have the same \textsl{CP} eigenvalue, one
has $\eta=+1$, otherwise $\eta=-1$.

\bigskip The three-body decay rate can be obtained from
eq.~\eqref{eqn:fermion_3body_width}, which can be easily generalized to also
account for neutrinos and quarks in the final state.  Note that in general one
has to add the matrix elements for the decays mediated by $\med=\widetilde
l_L$ and $\med=\widetilde l_R$, and compute the decay rate from the square of
the summed matrix elements. However, it turns out that all
`interference'-terms are suppressed by the bino--higgsino mixing,\footnote{
The reason is the following: If we consider a pure bino-slepton-lepton
interaction, the slepton $\widetilde l_L$ couples only to left-handed leptons
and the slepton  $\widetilde l_R$ only to right-handed ones. Thus both
channels are `orthogonal' in the limit where neutralino and slepton mixing are
neglected.}~which we neglect here.  Thus, it is possible to add the decay
rates directly.  Note that there is an additional contribution from a $Z^0$ on
the intermediate line~\cite{Bartl:1986hp}, which is subdominant for the
parameter range considered in ref.~\cite{Ibarra:2008kn}. Therefore, we also
neglect it here for simplicity. We assume, as above, degenerate sleptons.  The
decay rate summed over three generations of charged leptons and neutrinos is
thus (assuming $m_{\widetilde \nu}\simeq m_{\widetilde l}$)
\begin{eqnarray}
  \sum_{\ell,\nu} \Gamma(\fdm\rightarrow \ell\bar\ell\nn)  & \simeq &
  \frac{{g'}^4\theta^2}{64(2\pi)^3} \frac{m_{\fdm}^5}{24m_{\widetilde l}^4}
  \times 3 \times 18 \times \left( F_1 + 2\eta F_2 \right)\;.
\end{eqnarray}
In the hierarchical limit $m_{\nn} \ll m_{\fdm}$, the kinematical factor
approaches unity, $F_1 + 2\eta F_2 \rightarrow 1$. In the degenerate limit
$m_{\nn} \rightarrow m_{\fdm}$, one finds $F_1 + 2\eta F_2 \rightarrow
(2+\eta)(1-m_{\nn}^2/m_{\fdm}^2)^5/5$.  

\paragraph{}
The ratio of decays into $\gamma N$ to the decays into \textit{charged}
leptons is thus given by 
\begin{eqnarray}
  \frac{\Gamma(\fdm \rightarrow \gamma\nn)}{\sum_{\ell} \Gamma(\fdm
  \rightarrow \ell^+\ell^-\nn)} & \simeq & \frac{3\alpha_\text{em}}{8\pi}
  \underbrace{
  \frac{[3(1-4\eta)]^2}{51} \rule[-4mm]{0mm}{10mm} }_{\equiv R_\eta}
  \underbrace{\frac{ \left(1 - x\right)^3 \left(1 -
  \eta\sqrt{x}\right)^2  }{F_1(x) + 2\eta
  F_2(x)}}_{\equiv S_\eta}\;,
  \label{eqn:ratioHG}
\end{eqnarray}
with $x=m_{\nn}^2/m_{\fdm}^2$.  For the hidden gaugino, the decays
$\psi_\text{DM} \rightarrow Z^0 N$ and $\psi_\text{DM} \rightarrow h^0 N$ can
also be important since they occur at tree level. Their rates are given in
eqs. (A.1) and (A.3) of ref.~\cite{Ibarra:2008kn}.

According to eq.~\eqref{eqn:ratioHG}, the model-dependent factor $R_\eta$ is
here given by $R_+=1.6$ for $\eta=+1$ and $R_-=4.4$ for $\eta=-1$,
respectively, and hence of order one. The kinematical factor $S_\eta$ is
precisely of the form that was discussed in section
\ref{sec:scalar_intensity}, where we found that the two-body decay rate may
gain significantly in importance relative to the three-body decay rate if the
masses of the hidden gaugino $\psi_\text{DM}$ and the neutralino $N$ are
near-degenerate and the two particles have opposite \textsl{CP} parities.

\subsection{Intermediate vector: intensity of the gamma-ray line}
In the case of mediation by a vector, the ratio between two- and three-body
decay rates is
\begin{equation}
  \frac{\Gamma(\psi_\text{DM} \rightarrow N \gamma)}{\sum_\ell
  \Gamma(\psi_\text{DM} \rightarrow \ell^+ \ell^- N)} \simeq
  \frac{27\alpha_\text{em}}{8\pi} \frac{\left[\sum_\ell \left(\lambda_{\ell
  N}^L \lambda_{\ell \psi}^L - \eta \, \lambda_{\ell N}^R \lambda_{\ell
  \psi}^R\right)\right]^2 (1 - x)^3 (1 - \eta \, \sqrt{x})^2}{\sum_\ell (C_1^V
  F_1(x) + C_2^V F_2(x))},
\end{equation}
where $x \equiv m_N^2/m_{\psi_\text{DM}}^2$.  As before, it is useful to
consider the hierarchical limit $m_\nn/m_\fdm\to 0$, and the degenerate limit
$m_\nn/m_\fdm\to 1$, for which it is possible to capture the dependence on the
couplings in a factor $R_\eta^V$ and on kinematics in a model-independent
factor $S_\eta$,
\begin{align}\label{ratio-vector-RS}
  \frac{\Gamma(\psi_\text{DM} \rightarrow N \gamma)}{\sum_\ell
  \Gamma(\psi_\text{DM} \rightarrow \ell^+ \ell^- N)} \simeq \frac{27
  \alpha_\text{em}}{8 \pi} \times R_\eta^V(\lambda_{\ell N}^L,\lambda_{\ell
  \psi}^L,\lambda_{\ell N}^R,\lambda_{\ell \psi}^R) \times
  S_\eta(m_N/m_{\psi_\text{DM}}) \,.
\end{align}
The kinematical factors $S_\eta$ are identical to the case of scalar
mediation, see eq.~\eqref{S}. For the model-dependent factors, one finds
\begin{align}
  R^{V, {\rm hier}}_\eta(\lambda_{\ell N}^L,\lambda_{\ell
  \psi}^L,\lambda_{\ell N}^R,\lambda_{\ell \psi}^R) = \frac{\left[\sum_\ell
  \left(\lambda_{\ell N}^L \lambda_{\ell \psi}^L - \eta \, \lambda_{\ell N}^R
  \lambda_{\ell \psi}^R\right)\right]^2}{\sum_\ell \left[\big(\lambda_{\ell
  \psi}^{L2} +\lambda_{\ell \psi}^{R2} \big) \big(\lambda_{\ell N}^{L2}
  +\lambda_{\ell N}^{R2}\big) + 2\eta \, \lambda_{\ell \psi}^L \lambda_{\ell
  N}^L \lambda_{\ell \psi}^R \lambda_{\ell N}^R\right]}\;,
\end{align}
in the hierachical case, and
\begin{align}
  R^{V, {\rm deg}}_\eta(&\lambda_{\ell N}^L,\lambda_{\ell
  \psi}^L,\lambda_{\ell N}^R,\lambda_{\ell \psi}^R) = \\\nonumber&
  \frac{\left[\sum_\ell \left(\lambda_{\ell N}^L \lambda_{\ell \psi}^L -
  \eta\, \lambda_{\ell N}^R \lambda_{\ell \psi}^R\right)\right]^2 }{\sum_\ell
  \left[(\lambda_{\ell \psi}^{L2}\lambda_{\ell N}^{L2}+\lambda_{\ell
  \psi}^{R2}\lambda_{\ell N}^{R2})+\frac{2}{2+\eta}(\lambda_{\ell
  \psi}^{L2}\lambda_{\ell N}^{R2}+\lambda_{\ell \psi}^{R2}\lambda_{\ell
  N}^{L2})+ \frac{4\eta}{2+\eta}\lambda_{\ell \psi}^{L}\lambda_{\ell \psi}^{R}
  \lambda_{\ell N}^{L}\lambda_{\ell N}^{R}\right]}\;,
\end{align}
in the degenerate case. For purely chiral, e.g. left-handed, couplings one
finds that $R^{V, {\rm hier}}_\eta=R^{V, {\rm deg}}_\eta=R^{\rm chir}$
coincides with the expression~\eqref{Rchir} for the scalar case, as does the
kinematical factor. For a generic set of couplings, $R_\eta^V$ is roughly of
order one.

\begin{figure}
  \begin{center}
    \includegraphics[width=12cm]{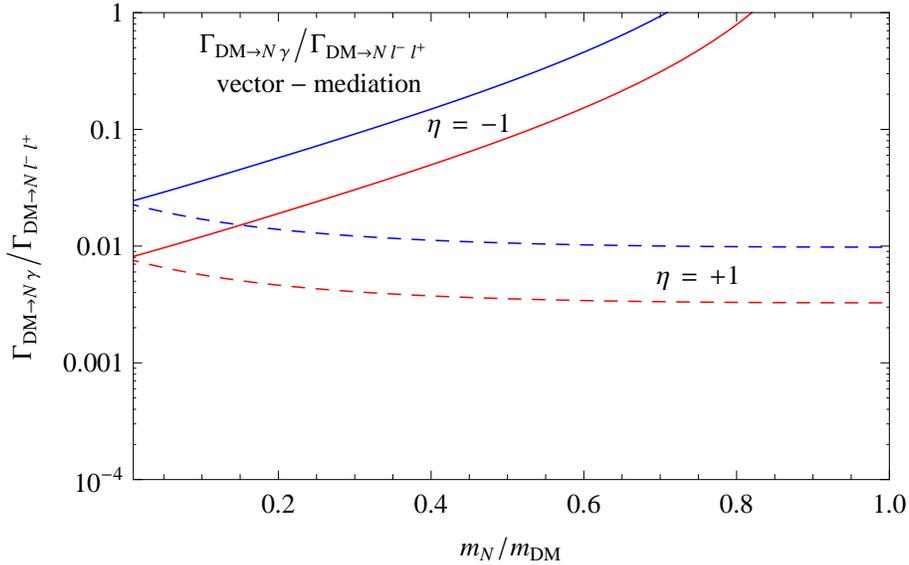}
  \end{center}
  \caption{Ratio of the decay rates $\Gamma(\psi_\text{DM} \rightarrow \gamma
  N)/\sum_\ell \Gamma(\psi_\text{DM} \rightarrow \ell^+ \ell^- N)$ for
  decay mediated by a heavy vector. Otherwise, the four cases are identical to
  the ones shown in fig.~\ref{fig:ratio}.}\label{fig:ratioVector}
\end{figure}

Note that the prefactor of the ratio of decay rates,
eq.~\eqref{ratio-vector-RS}, for mediation by a vector is larger by a factor
of nine compared to mediation by a scalar, eq.~\eqref{ratio-scalar-RS}.  Thus,
in the hierarchical case $m_\nn/m_\fdm\to 0$ as well as in the degenerate case
with $\eta=+1$ one finds a ratio between two-body and tree-level decay of the
order of $10^{-2}$, one order of magnitude larger than for the scalar case. In
addition, when $\eta=-1$ the gamma-ray line is further enhanced for
$m_\nn/m_\fdm\to 1$ by the kinematic effect discussed in
section~\ref{sec:scalar_intensity}.  The ratio for some specific examples is
shown in fig.~\ref{fig:ratioVector}.

Therefore, there are scenarios with dark matter decay mediated by heavy
vectors where a gamma-ray line can be fairly intense, despite being
loop-suppressed, while at the same time being in agreement with the
electron/positron measurements.

\section{Radiative decay of scalar dark matter}\label{sec:scalarDM}

\begin{figure}
  \begin{center}
    \includegraphics{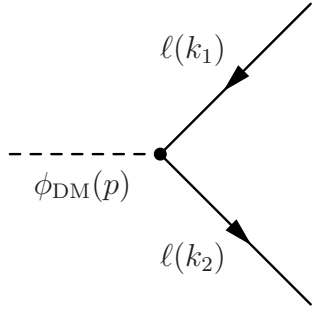}\\
    \caption{Tree-level decay of scalar dark matter.}
    \label{scalar_tree}
  \end{center}
\end{figure}

\begin{figure}
  \begin{center}
    \includegraphics{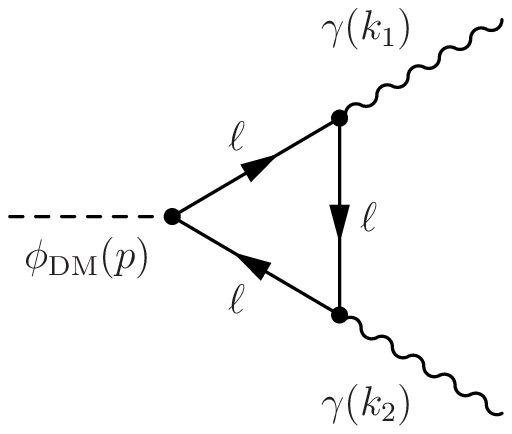}~~~~~
    \includegraphics{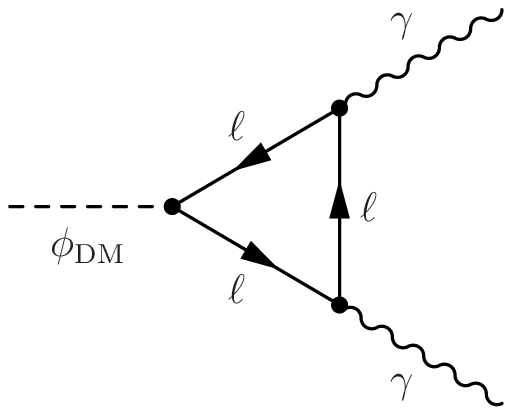}\\
    \caption{Diagrams contributing to the two-body decay of scalar dark matter
    into two photons at the one-loop level.}
    \label{scalar_loop}
  \end{center}
\end{figure}

We now consider the case that the dark matter particle is a (pseudo-)scalar
which we denote by $\phi_\text{DM}$. In this case, the symmetries allow for
the decay into a pair of charged leptons at tree level,
$\phi_\text{DM}\rightarrow \ell^+ \ell^-$. We describe this by an effective
Lagrangian that describes a direct interaction between dark matter and charged
leptons,
\begin{equation}\label{eqn:scalar_lagrangian}
  \mathcal{L}_\text{eff} = -\bar{\ell} \left[\lambda_{\ell \phi}^L P_L +
  \lambda_{\ell \phi}^R P_R\right] \ell \, \phi_\text{DM} + \text{h.c.}\;.
\end{equation}
If the dark matter particle is a parity eigenstate, one has $\lambda_{\ell
\phi}^L = \lambda_{\ell \phi}^R$ for a scalar and $\lambda_{\ell \phi}^L =
-\lambda_{\ell \phi}^R$ for a pseudo-scalar.

\subsection{The decay \boldmath$\phi_\text{DM} \rightarrow \ell^+ \ell^-$}
The effective Lagrangian (\ref{eqn:scalar_lagrangian}) will give rise to the
tree-level decay shown in fig.~\ref{scalar_tree}. The corresponding decay
width is
\begin{equation}
  \Gamma\left(\phi_\text{DM} \rightarrow \ell^+ \ell^-\right) = \frac{1}{16
  \pi \, m_{\phi_\text{DM}}} \left|\mathcal{M}\right|^2 \sqrt{1 - \frac{4
  m_\ell^2}{m_{\phi_\text{DM}}^2}},
\end{equation}
where $m_{\phi_\text{DM}}$ and $m_\ell$ are the mass of the dark matter and
the charged leptons, respectively, and the amplitude is given by
\begin{equation}
  \left|\mathcal{M}\right|^2 = m_{\phi_\text{DM}}^2 \left(\left|\lambda_{\ell
  \phi}^L\right|^2 + \left|\lambda_{\ell \phi}^R\right|^2\right) - 2 m_\ell^2
  \left|\lambda_{\ell \phi}^L + \lambda_{\ell \phi}^R\right|^2\;.
\end{equation}
Thus, in the case of equal left- and right-handed couplings, $\lambda_{\ell
\phi}^L = \lambda_{\ell \phi}^R \equiv \lambda_{\ell \phi}$, one gets
\begin{equation}
  \Gamma\left(\phi_\text{DM} \rightarrow \ell^+ \ell^-\right) =
  \frac{|\lambda_{\ell \phi}|^2}{8 \pi} m_{\phi_\text{DM}} \left(1 - \frac{4
  m_\ell^2}{m_{\phi_\text{DM}}^2}\right)^{3/2}\;.
\end{equation}
For $m_\ell \ll m_{\phi_\text{DM}}$ this corresponds to a lifetime
\begin{equation}
  \tau_{\phi_\text{DM} \rightarrow \ell^+ \ell^-} \simeq 2 \times
  10^{26}\,\text{s}\,\left(\frac{10^{-26}}{\left|\lambda_{\ell
  \phi}\right|}\right)^2
  \left(\frac{1\,\text{TeV}}{m_{\phi_\text{DM}}}\right)\;.
\end{equation}

\subsection{The decay \boldmath$\phi_\text{DM} \rightarrow \gamma \gamma$}
By combining the external lepton lines into a loop, decays into two
monochromatic photons radiated off the charged lepton loop are induced at the
quantum level (see fig.~\ref{scalar_loop}).

For equal left- and right-handed couplings, $\lambda_{\ell \phi}^L =
\lambda_{\ell \phi}^R \equiv \lambda_{\ell \phi}$, and in the limit $m_\ell
\ll m_{\phi_\text{DM}}$ (see app. \ref{app:scalar} and
\cite{Resnick:1973vg,Spira:1995rr}),
\begin{equation}
  \Gamma(\phi_\text{DM} \rightarrow \gamma \gamma) =
  \frac{m_{\phi_\text{DM}}^3}{16 \pi} \left(\frac{e^2}{16 \pi^2}\right)^2
  \left|\sum_\ell \frac{\lambda_{\ell \phi}}{m_\ell} A_f(\tau_\ell)\right|^2,
\end{equation}
where for the relevant limit $\tau \gg 1$ one has
\begin{equation}
  A_f(\tau) \simeq \frac{1}{\tau} \left\{2 - \frac{1}{2} \left(\ln(4\tau) -
  i\pi\right)\right\}.
\end{equation}
Thus, when taking only one lepton species into account, we obtain for the
ratio between the decay into photons and charged leptons
\begin{align}
  \frac{\Gamma(\phi_\text{DM} \rightarrow \gamma
  \gamma)}{\Gamma(\phi_\text{DM} \rightarrow \ell^+ \ell^-)} &\simeq
  \frac{\alpha_\text{em}^2}{2\pi^2} \frac{m_\ell^2}{m_{\phi_\text{DM}}^2}
  \left|2 - \frac{1}{2} \left(\ln(4\tau_\ell) - i\pi\right)^2\right|^2
  \nonumber\\
  &\simeq 10^{-9} \left(\frac{m_\ell}{106~\text{MeV}}\right)^2
  \left(\frac{1~\text{TeV}}{m_{\phi_\text{DM}}}\right)^2\;.
\end{align}
We see that for scalar dark matter, the decay into two photons is highly
suppressed by the factor $m_\ell^2/m_{\phi_\text{DM}}^2$ compared to the decay
into a pair of charged leptons. In addition to this helicity-suppression
factor there appears a factor $\alpha_\text{em}^2/\pi^2$ as opposed to
$\alpha_\text{em}/\pi$ for fermionic dark matter, since the loop contains two
photon vertices, and both the tree-level and one-loop decays are two-body
decays. The same suppression factors occur for pseudo-scalar dark matter and
for the decay into massive gauge bosons. Thus, there appears to be no hope of
detecting a gamma-ray line in this case. For more general expressions for the
decay rates, see app. \ref{app:scalar}.

\section{Observational constraints}\label{sec:observational_constraints}
The observation of a cosmic gamma-ray line at TeV energies would be a strong
hint for the dark matter interpretation of the PAMELA/Fermi LAT $e^\pm$
anomalies. On the other hand, the non-observation of gamma-ray lines can be
used to constrain the above leptophilic models, which induce these lines at
one loop, as discussed above. The gamma-ray lines that originate from dark
matter decay inside the Milky Way halo could be observed in the isotropic
diffuse gamma-ray flux. Furthermore, lines may be observable in the flux from
nearby galaxies and galaxy clusters.

At intermediate energies, satellite instruments such as Fermi LAT are a very
sensitive probe for gamma-ray lines in the Galactic flux. At higher energies,
Imaging Atmospheric Cherenkov Telescopes (IACTs) provide important
information. For the future, the proposed Cherenkov Telescope Array (CTA) is
expected to improve the flux sensitivity of current IACTs (MAGIC, H.E.S.S.,
VERITAS) by an order of magnitude. We put some emphasis on IACTs, since these
instruments are capable of probing the high energy ranges relevant to the dark
matter interpretation of PAMELA/Fermi LAT.

\subsection{Fermi LAT}
The flux of monochromatic gamma rays from the decay of dark matter in the
Milky Way halo is given by a line-of-sight integral over the dark matter
distribution~\cite{Bertone:2007aw}. This component of the gamma-ray flux is
explicitly given by
\begin{equation}
  \frac{dJ_\text{dm}^\text{halo}}{dE} = \frac{\Gamma(\psi_\text{DM} \to \gamma
  N)}{4 \pi m_{\psi_\text{DM}}}\, \delta\left(E_\gamma - E\right)
  \int_\text{l.o.s.} d\vec{l} \, \rho_\text{DM}^\text{MW}(\vec{l})\;,
  \label{eqn:JdmHalo}
\end{equation}
where $\Gamma(\psi_\text{DM} \to \gamma N)$ denotes the partial decay width of
dark matter particles for two-body decays involving a photon and a neutral
particle $N$. When the neutral particle is massless, we will write $\nu$
instead of $N$ in the following. Furthermore, $m_{\psi_\text{DM}}$ is the mass
of the dark matter particle, $E_\gamma$ is the energy of the produced
gamma-ray line as given by eq.~\eqref{eqn:E_gamma}, while
$\rho_\text{DM}^\text{MW}$ is the Milky Way's dark matter halo density
profile. We adopt the NFW profile here, which has the form
\begin{equation}
  \rho_\text{DM}^\text{MW}(r) = \frac{\rho_\text{c}}{r/r_\text{c} \left(1 + r/r_\text{c}\right)^2}\;,
  \label{eqn:NFWprofile}
\end{equation}
with the parameters $\rho_\text{c}=0.35\GeV\cm^{-3}$ and $r_\text{c}=20$
kpc~\cite{Navarro:1996gj, Abdo:2010nc}, leading to a local dark matter density
of 0.4 GeV/cm$^3$ \cite{Catena:2009mf}. The gamma-ray flux from dark matter
decay inside the Galactic halo has only a mild angular dependence and can be
considered as isotropic for our purposes (for details on anisotropies in the
Galactic gamma-ray flux from dark matter decay, see
refs.~\cite{Bertone:2007aw, Ibarra:2009nw}).  The extragalactic contribution
stemming from the decay of dark matter at cosmological distances is generally
fainter than the Galactic flux, and we will neglect this component here.

The Fermi LAT collaboration has conducted a negative search for Galactic
gamma-ray lines in the diffuse flux in the energy range from 30 to 200
GeV~\cite{Abdo:2010nc}. For the halo profile~\eqref{eqn:NFWprofile}, we plot
the resulting 2$\sigma$ limits on the partial decay width corresponding to
$\psi_\text{DM}\to \gamma \nu$ in fig.~\ref{fig:lifetime_constraints}.  Most
interestingly, the Fermi LAT observations can constrain the dark matter decay
into photons at the one-loop level if the total dark matter lifetime is of the
order $10^{26}$ seconds. Thus, the Fermi LAT bounds on gamma-ray lines can be
relevant for dark matter scenarios with $m_{\psi_\text{DM}}\simeq300$ -- $400$
GeV, which can provide a possible explanation for the rise in the positron
fraction observed by PAMELA (see, e.g. \cite{Ibarra:2008jk}).

\subsection{Imaging Atmospheric Cherenkov Telescopes}
IACTs are important tools to constrain scenarios with dark matter masses in
the multi-TeV range. One property of these instruments is that the atmospheric
showers induced by cosmic-ray electrons or gamma rays cannot be distinguished
easily, since both particle species initiate similar electro-magnetic cascades
in the atmosphere. The large cosmic-ray electron flux hence comprises an
irreducible background for high energy gamma-ray observations. Since the
electron background is expected to be very isotropic --- in contrast to the
gamma rays --- it can be removed by calculating differences between fluxes
that are observed in different neighboring regions of the sky. As a result,
IACTs are best suited to observe localized sources, whereas diffuse signals
such as those resulting from dark matter decay are more difficult to discern
from the background unless they exhibit sharp spectral features. Constraints
on the gamma-rays from decaying dark matter can be derived in two different
ways.  First, one can observe point-like sources like M31. Second, by using
the observed electron+gamma-ray flux (potentially also contaminated by
unrejected protons), one can derive upper limits on the Galactic halo signal
from dark matter decay. If the statistics are good enough, one could even hope
to see spectral features in the electron+gamma-ray flux, or translate their
non-observation into bounds on the corresponding dark matter decay width. This
will be described in the context of the CTA below.\medskip

\begin{figure}
  \begin{center}
    \includegraphics[width=0.7\linewidth]{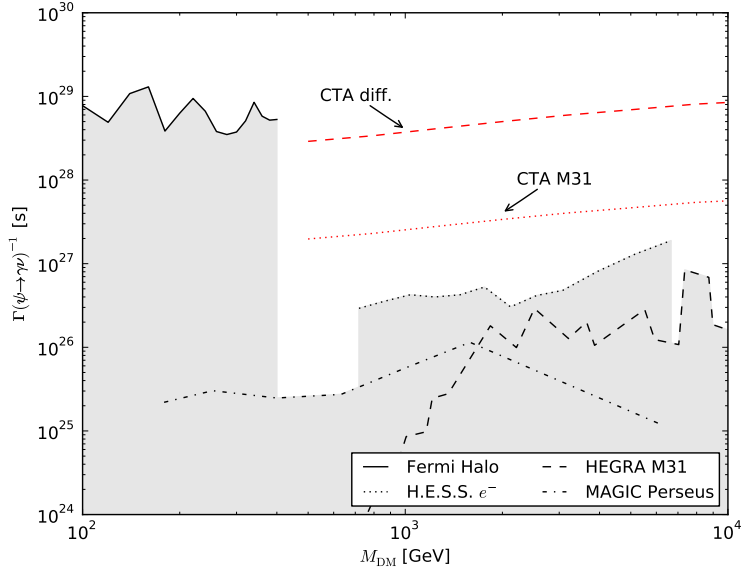}
    \caption{Lower bounds on the inverse decay width of dark matter decaying
    into gamma-ray lines via $\psi_\text{DM}\to \gamma \nu$ are shown as
    black lines. The bounds on this decay channel come from line searches in
    M31 by HEGRA, from line searches in the diffuse flux by Fermi LAT and from
    observations of the Perseus cluster by MAGIC. Further bounds can be
    derived from the $(\gamma+)e^-$ observations of H.E.S.S. Our estimates of
    the reach of the future CTA in measurements of the flux from M31 or
    spectral variations in the diffuse $\gamma+e^-$ flux are shown as red
    lines.}
    \label{fig:lifetime_constraints}
  \end{center}
\end{figure}

The HEGRA collaboration has published constraints on the gamma-ray line flux
from M31~\cite{Aharonian:2003xh}. These bounds can be converted into 99\%
C.L.~limits on the decay width of dark matter into gamma-ray lines. HEGRA
observed a region with an opening-angle of $\theta_\text{obs} = 0.105^\circ$,
corresponding to the inner 1.4 kpc region of M31. The expected flux of gamma
rays from dark matter decay from M31 can be derived as follows. We define
$\theta$ to be the angle between the line-of-sight and the ray that passes
through our position and the center of M31. Each angle $\theta$ then
corresponds to an `impact parameter' $R$.  If $D$ is the distance to the
target ($D=770$ kpc in case of M31), we have $R \simeq D\,\theta$. The
gamma-ray flux from dark matter decay in M31 within the opening angle
$\theta_\text{obs}$ is then
\begin{equation}
  \frac{dJ^\text{M31}_\text{DM}}{dE}=
  \frac{\Gamma(\psi\to\gamma N)}{4 \pi m_\text{DM}} \,\delta(E_\gamma-E) \, 2
  \pi \int_0^{\theta_\text{obs}} d\theta \, \sin \theta
  \int_{-\infty}^{\infty} ds \, \rho_\text{DM}^\text{M31}(\sqrt{s^2 + R^2})\;,
  \label{eqn:M31flux}
\end{equation}
where the first integral is over the solid-angle, whereas the second integral
is over the line-of-sight. For the dark matter density profile of M31 we adopt
the NFW profile with values given in ref.~\cite{Boyarsky:2007ay},
$\rho_\text{c}=2.0 \GeV/\cm^3$ and $r_\text{c} = 8.31$ kpc. The other profiles
from ref.~\cite{Boyarsky:2007ay} lead to similar constraints.  The signal from
decaying dark matter has a relatively large angular extent due to the linear
dependence on the halo profile, and can leak into the off-region which is used
to estimate the background fluxes of the IACT. The details of this effect
depend on the details of the adopted off-region and are different for each
observation. Here and below, we incorporate this effect simply by subtracting
from eq.~\eqref{eqn:M31flux} a flux corresponding to the dark matter-induced
flux emitted at $\theta=2\,\theta_\text{obs}$. This should lead to correct
bounds within a factor of two. Our results are shown in
fig.~\ref{fig:lifetime_constraints}.

Upper limits on the gamma-ray flux from the Perseus galaxy cluster were
presented by the MAGIC collaboration in ref.~\cite{Aleksic:2009ir}. For the
density profile of the Perseus cluster we take the NFW profile with
$r_\text{c}=384$ kpc and $\rho_\text{c}=0.04\GeV\cm^{-3}$, the obsevational
angle is $\theta_\text{obs}=0.15^\circ$, and the distance to the Perseus
cluster is 78~Mpc. The resulting 95\% C.L. bounds are shown in
fig.~\ref{fig:lifetime_constraints}.\footnote{We take the limits corresponding
to $\Gamma=-2.5$ from tab.~4 of ref.~\cite{Aleksic:2009ir}.} Since the energy
threshold of the MAGIC telescope is very low, we can constrain gamma-ray lines
with energies down to $100\GeV$.\bigskip

\begin{figure}
  \begin{center}
    \includegraphics[width=0.7\linewidth]{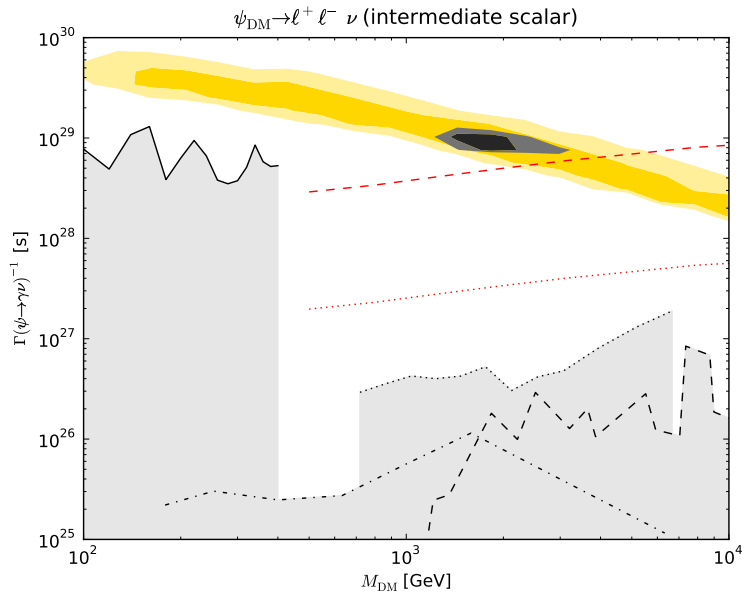}
    \caption{Same as fig.~\ref{fig:lifetime_constraints}. In addition, the
    orange and gray shaded regions show the parts of the parameter space that
    are relevant for the dark matter explanation of the PAMELA/Fermi $e^\pm$
    anomalies with the flavor-democratic decay channel
    $\psi_\text{DM}\to\ell^+\ell^-\nu$.  The intermediate particle is assumed
    to be a scalar, in which case the branching ratio into monochromatic
    photons can be as large as $\text{BR}(\psi_\text{DM}\to\gamma\nu)
    \simeq3\times3\alpha_\text{em}/(8\pi)$, which we assume here.}
    \label{fig:bounds_leplepnu}
  \end{center}
\end{figure}

\begin{figure}
  \begin{center}
    \includegraphics[width=0.7\linewidth]{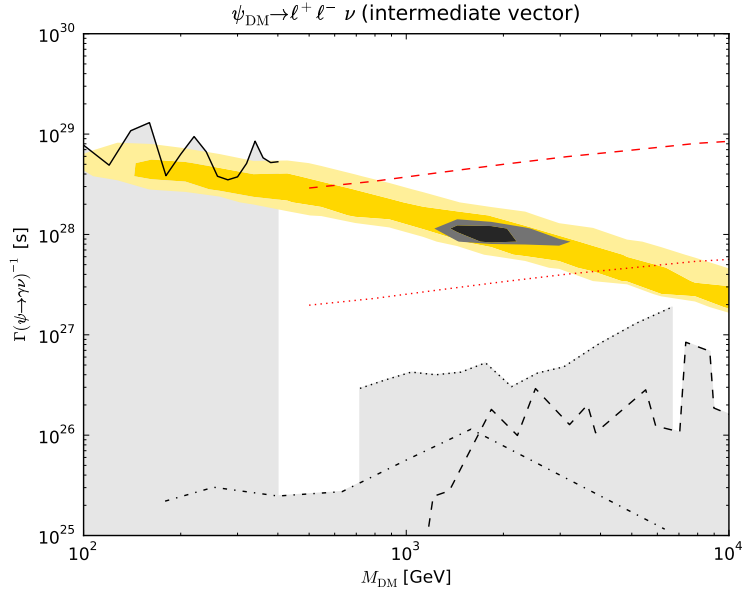}
    \caption{Same as fig.~\ref{fig:bounds_leplepnu}, but assuming that the
    intermediate particle is a vector, in which case the branching ratio into
    monochromatic photons can be as large as
    $\text{BR}(\psi_\text{DM}\to\gamma\nu)
    \simeq3\times27\alpha_\text{em}/(8\pi)$, which we assume here.}
    \label{fig:bounds_leplepnu_vector}
  \end{center}
\end{figure}

\begin{figure}
  \begin{center}
    \includegraphics[width=0.7\linewidth]{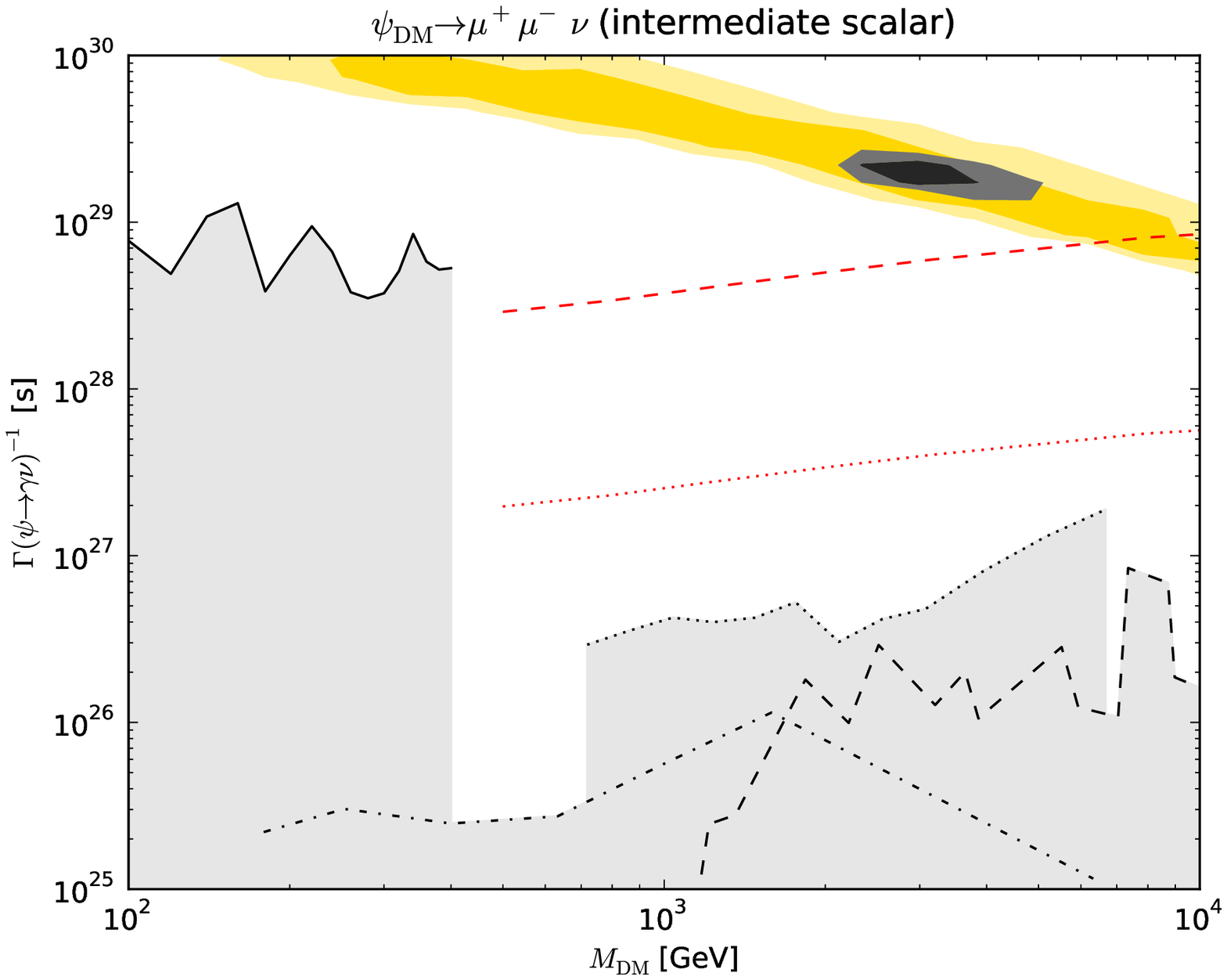}
    \caption{Same as fig.~\ref{fig:bounds_leplepnu}, but for decay into
    $\psi_\text{DM}\to\mu^+\mu^-\nu$. The intermediate particle is assumed to
    be a scalar, leading to a maximal branching ratio of
    $\text{BR}(\psi_\text{DM}\to\gamma\nu) \simeq3\alpha_\text{em}/(8\pi)$,
    which we assume here.}
    \label{fig:bounds_mumunu}
  \end{center}
\end{figure}

\begin{figure}
  \begin{center}
    \includegraphics[width=0.7\linewidth]{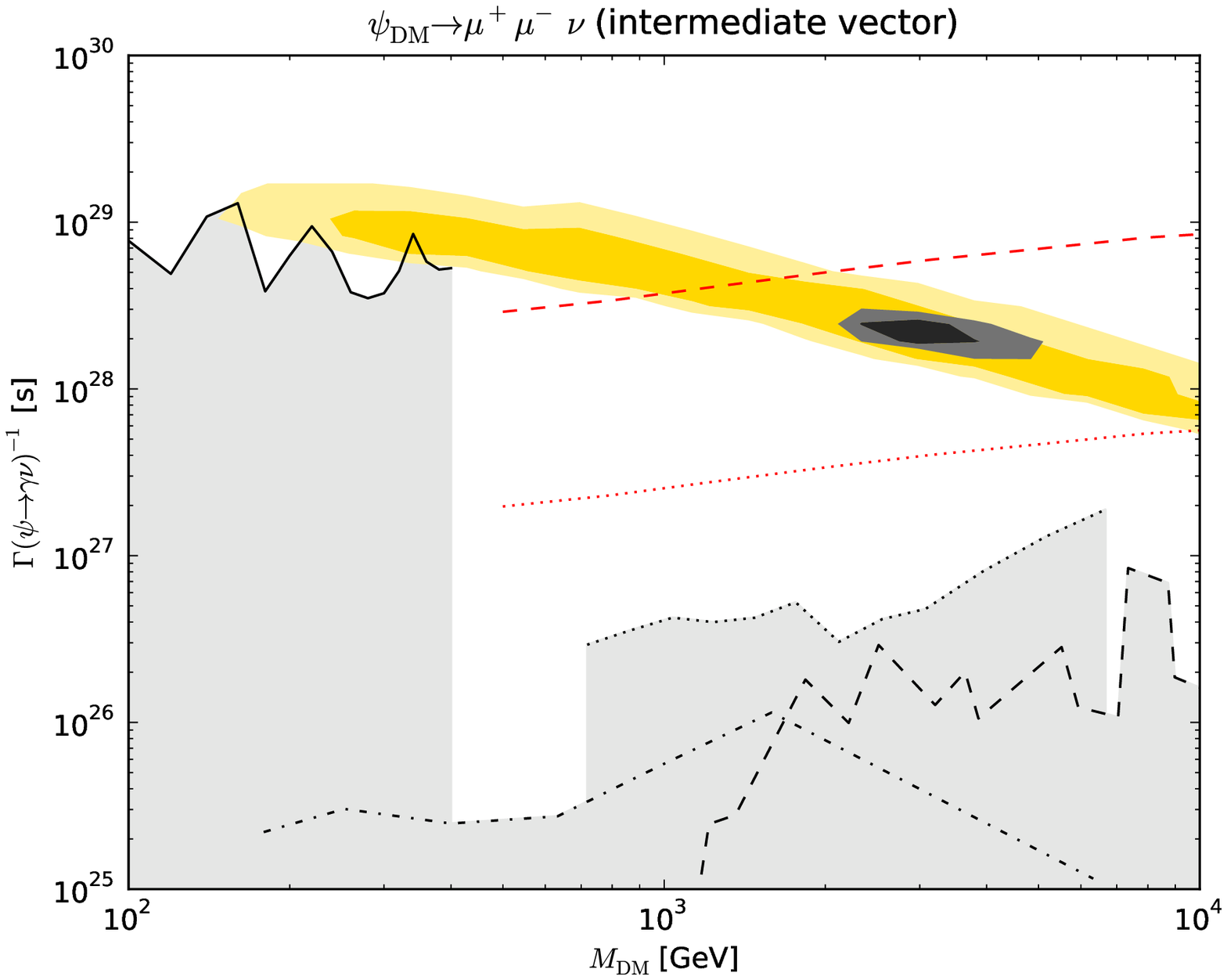}
    \caption{Same as fig.~\ref{fig:bounds_leplepnu}, but for decay into
    $\psi_\text{DM}\to\mu^+\mu^-\nu$. The intermediate particle is assumed to
    be a vector, leading to a maximal branching ratio of
    $\text{BR}(\psi_\text{DM}\to\gamma\nu) \simeq27\alpha_\text{em}/(8\pi)$,
    which we assume here.}
    \label{fig:bounds_mumunu_vector}
  \end{center}
\end{figure}

\begin{figure}
  \begin{center}
    \includegraphics[width=0.7\linewidth]{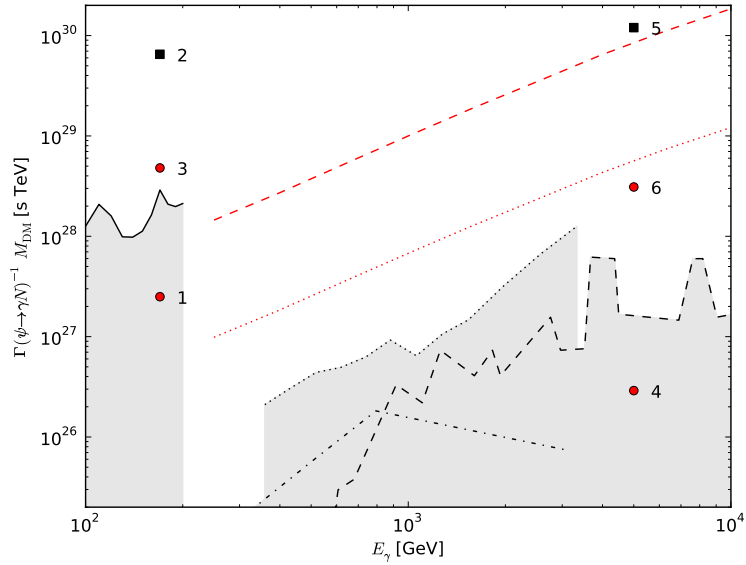}
    \caption{Same as fig.~\ref{fig:lifetime_constraints}, but with different
    scaling of the axes to allow for non-vanishing $m_N$. The black squares
    and red dots show the predictions for the different benchmark scenarios
    summarized in tab.~\ref{tab:benchmarks}. Black squares correspond to
    scenarios with $\eta=+1$, while red dots correspond to $\eta=-1$. The last
    benchmark point in tab.~\ref{tab:benchmarks} lies outside of the shown
    parameter region.}
    \label{fig:bounds_benchmarks}
  \end{center}
\end{figure}

\begin{table}
  \centering
  \begin{tabular}{cccccccc}
    \#&Channel & $\eta$ & $m_\text{DM}$ [GeV] & $E_\gamma$ [GeV] & 
    $m_N$ [GeV] & $\Gamma_{\ell^+\ell^-N}^{-1}$ [s] &
    $\frac{m_\text{DM}}{\Gamma_{\psi\to \gamma N}}$ [s\ TeV]\\\hline
    1&$e_L^-e_L^+N$   &$-1$ &1000   &170  &812.4 &$2.5\times10^{26}$
    &$2.47\times10^{27}$
    \\
    2&$e^-_Le^+_LN$   &$+1$ &500    &170  &282.8 &$5\times10^{26}$
    &$6.51\times10^{29}$\\
    3&$e^-_Le^+_LN$  &$-1$ &400    &170  &154.9 &$6.3\times10^{26}$
    &$4.83\times10^{28}$\\
    4&$\mu^-_L\mu^+_LN$ &$-1$ &100000 &5000 &94868 &$4.5\times10^{24}$
    &$2.87\times10^{26}$\\
    5&$\mu^-_L\mu^+_LN$ &$+1$ &15000  &5000 &8660  &$3\times10^{25}$
    &$1.18\times10^{30}$\\
    6&$\mu^-_L\mu^+_LN$ &$-1$ &15000  &5000 &8660  &$3\times10^{25}$
    &$3.06\times10^{28}$\\
    7&$\mu_L\mu_RN$ &$+1$ &15000  &5000 &8660  &$3\times10^{25}$
    & $3.42\times10^{34}$
  \end{tabular}
  \caption{Benchmark scenarios. In the first three cases, the three-body decay
  produces only electrons. In the last four cases, the three-body decay
  produces muons. The gamma-ray line intensity of these scenarios is
  illustrated in fig.~\ref{fig:bounds_benchmarks}.}
  \label{tab:benchmarks}
\end{table}

The H.E.S.S.~collaboration has published measurements of the electron flux at
TeV energies~\cite{Aharonian:2008aa, Aharonian:2009ah}. The measured electron
flux may be contaminated with diffuse gamma rays by no more than
$\approx50\%$~\cite{Aharonian:2008aa}. This fact allows to translate the
electron flux into upper bounds on gamma-ray lines from dark matter decay in
the Galactic halo. For energies above 1 TeV, we derived $2\sigma$-bounds from
the fluxes shown in fig.~3 of ref.~\cite{Aharonian:2008aa}. For energies below
1 TeV, where the H.E.S.S.~results overlap with the Fermi LAT measurements of
the electron flux, $1\sigma$-upper limits on the amount of diffuse gamma rays
were derived by comparing the H.E.S.S.~and the Fermi LAT electron fluxes in
ref.~\cite{Meyer:2009cz}. These upper limits can also be used as bounds on
gamma-ray lines. Our results are shown in
fig.~\ref{fig:lifetime_constraints}.\bigskip

\paragraph{Prospects for the CTA.} We will now briefly discuss observational
prospects for the future Cherenkov Telescope Array (CTA, see
ref.~\cite{Consortium:2010bc} for a recent discussion). The expected
2$\sigma$-limit from M31 that the CTA could produce can be roughly estimated
by
\begin{equation}
  \langle\Jdm^\text{M31}\rangle_\text{on}\lesssim
  \frac{\max(2\sqrt{\Non},3.1)}{T\Aeff}\;,
  \label{eqn:M31bound}
\end{equation}
where $N_\text{on}$ denotes the number of measured events in the on-region,
$T$ is the measurement time, $A_\text{eff}$ denotes the effective area of the
instrument (we take $A_\text{eff} \approx 2\,\text{km}^2$ at 5 TeV, and let it
scale with the energy as in ref.~\cite{delaCallePerez:2006gc}, fig.~17a), and
$\langle\Jdm^\text{M31}\rangle_\text{on}$ is the gamma-ray flux from M31
averaged over the on-region. As on-region, we take a circle with $1.0^\circ$
radius around the center of M31, and for the off-region we assume that the
solid-angle of the off-region is much larger than the on-region, $\Ooff \gg
\Oon$. If we assume that only the background is observed, and no signal is
coming from M31, $\Non$ can be estimated by
\begin{align}
  \label{eqn:Nexp}
  \bar N_\text{on}=\Oon T \Aeff (J_{e^-} + \epsilon_r J_p)\;.
\end{align}
Here, $J_{e^-}$ and $J_p$ denote the cosmic-ray electron and proton fluxes,
respectively, and $\epsilon_r$ is the rejection factor of protons.  Fluxes
have to be integrated over an energy range that corresponds to the energy
resolution of the detector (around 10\%, taken from
ref.~\cite{Consortium:2010bc}, fig.~23, scenario E). For the cosmic-ray
electron and proton fluxes at high energies we take
\begin{equation}
  \frac{dJ_{e^-}}{dE}=1.17\times10^{-11}\left( \frac{E}{\TeV}
  \right)^{-3.9}
  \text{cm}^{-2}\s^{-1}\sr^{-1}\GeV^{-1}\;,
  \label{eqn:Jee}
\end{equation}
from ref.~\cite{Aharonian:2008aa}, and
\begin{equation}
  \frac{dJ_p}{dE} = 8.73 \times 10^{-9}
  \left(\frac{E}{\text{TeV}}\right)^{-2.7} \cm^{-2}\s^{-1}\sr^{-1}\GeV^{-1}\;,
\end{equation}
from ref.~\cite{delaCallePerez:2006gc}, respectively, in agreement with the
cosmic-ray measurements. Furthermore, at energies below 1 TeV, the electron
flux becomes somewhat harder with a spectral index of $\simeq-3.1$, and we
replace eq.~\eqref{eqn:Jee} with a flux fitting the results of
ref.~\cite{Abdo:2009zk} in this energy regime. Taking into account also other
cosmic-ray species beside the protons would only have minor impact on our
results. The proton rejection factor is set to
$\epsilon_r\approx10^{-2}$~\cite{Consortium:2010bc, delaCallePerez:2006gc},
and we use as observational time of M31 $T=20\,\text{h}$. Our estimates for
the limits that the CTA could produce from M31 observations in the future are
shown in fig.~\ref{fig:lifetime_constraints} by the lower dashed line. They
are almost two orders of magnitude better than the limits derived from the
HEGRA observation. This is mainly due to the increased effective area of CTA,
but also due to the larger on-region that we adopted in our estimates. For
decaying dark matter it is not optimal to search for point-source signals, as
was done in the HEGRA analysis, for example. Using a larger on-region
typically leads to better results.

Instead of using the spatial variations of the observed cosmic-ray flux to
derive constraints on dark matter decay in extragalactic sources, one can also
derive constraints from the non-observation of spectral line features in the
diffuse flux, which could come from dark matter decaying into gamma-ray lines
in the Galactic halo. In this case it is best to consider data from large
fractions of the sky, to maximize the statistics. The expected ``halo''-bound
that the CTA will presumably reach then follows from
\begin{equation}
  \langle\Jdm^\text{halo}\rangle_\text{sky}\leq
  \frac{2\sqrt{N}}{T\Aeff\Omega}\;,
  \label{eqn:HaloBoundDef}
\end{equation}
where $\langle\Jdm^\text{halo}\rangle_\text{sky}$ denotes the gamma-ray flux
coming from dark matter decaying in our Galactic halo, averaged over all
angles. We assume that the data will be good enough to estimate the background
by fitting a power-law to the observed flux at energies close to the line,
similar to the analysis in ref.~\cite{Abdo:2010nc}, and we neglect the
statistical uncertainties in the background estimate.\footnote{Note that this
is different from our treatment of the H.E.S.S.~electron flux, where we only
required that the predicted line signal is below the observed fluxes, without
any attempt to subtract a power-law background.} In
eq.~\eqref{eqn:HaloBoundDef}, $N$ is the total number of observed events,
including electrons, gamma-rays and protons that pass the cuts. The region
$\Omega$ is taken to be as large as possible to maximize the statistics (we
assume $\Omega=\pi(3^\circ)^2$), and as observational time we take $T=1000$ h.
As above, we integrate over energy bands which correspond to the anticipated
energy resolution of CTA. Our resulting estimates for the bounds that CTA
could obtain observing the diffuse flux are shown in
fig.~\ref{fig:lifetime_constraints} by the upper dashed line. As can be seen
from this figure, the bounds on gamma-ray lines from dark matter decay that
can be put by looking at spectral variations in the observed diffuse fluxes
can be even stronger than the ones that can be derived from flux limits on
point-like sources like M31.

\subsection{Discussion}
\paragraph{The case \boldmath$m_N \rightarrow 0$.}
We first discuss the case where dark matter decays into a photon and a
massless particle. In fig.~\ref{fig:lifetime_constraints} we present a
collection of the lower bounds on the inverse decay width for two-body decays
into a monochromatic photon and a massless particle as determined by the
methods described in the previous subsection.  For dark matter masses between
100 and 400 GeV, the line searches in the diffuse Galactic flux by the Fermi
LAT constitute the strongest constraints.  At higher energies, Cherenkov
telescopes provide important information. As far as constraints from
particular sources are concerned, we show the constraints from HEGRA
observations of M31 and MAGIC observations of the Perseus cluster. We also
plot the constraints from the diffuse electron flux observed by H.E.S.S.
Lastly, we show our estimates for the reach of the future CTA which could
improve current limits by almost two orders of magnitude at energies above a
few hundred GeV.

In fig.~\ref{fig:bounds_leplepnu} we show the same constraints together with
shaded regions indicating the part of the parameter space relevant to PAMELA
and Fermi for the gamma-ray lines induced by the decay $\psi_\text{DM}
\rightarrow \ell^+ \ell^- \nu$. The orange regions correspond to the fit to
the positron fraction as measured by PAMELA, whereas the dark gray regions
correspond to the fit to the total $e^\pm$ flux as measured by Fermi LAT. In
both cases, the lighter shades indicate the $5\sigma$ confidence level around
the best-fit point, while the darker shades indicates the $3\sigma$ confidence
level. We only regard the data points above 10 GeV, which are not
significantly affected by solar modulation. For the background fluxes of
secondary electrons and positrons, we assume the `model 0'
backgrounds~\cite{Grasso:2009ma} as parametrized in~\cite{Ibarra:2009dr}. In
the energy range of interest, we assume that the primary electron flux is
given by a simple power law. At each point in the
$(m_\text{DM},\tau_\text{DM})$-plane, we then allow the power-law index of the
primary electron flux to vary between $-3.0$ and $-3.3$, whereas the
normalization is fitted to the data. We find that the relevant parameter space
is not constrained by current instruments, but could be constrained by CTA in
the future.

The same plot is shown in fig.~\ref{fig:bounds_leplepnu_vector}, but assuming
that the decay is mediated by an intermediate vector particle, in which case
the branching ratio can be as large as $3\times 27\alpha_\text{em}/(8\pi)$,
which we assume in the figure. This is about an order of magnitude larger than
in the case of mediation by a scalar, and one can see that in this case the
CTA can indeed constrain a significant part of the parameter space relevant to
the dark matter interpretations of PAMELA and Fermi. Analogously, in
figs.~\ref{fig:bounds_mumunu},~\ref{fig:bounds_mumunu_vector} we show the
corresponding plots for the lines induced by the decay $\psi_\text{DM}
\rightarrow \mu^+ \mu^- \nu$, in the cases of an intermediate scalar and an
intermediate vector particle, respectively. In these scenarios the expected
line signal is somewhat weaker.

\paragraph{The case \boldmath$m_N \sim m_{\psi_\text{DM}}$.}
Let us now turn to the case where the mass $m_\nn$ of the neutral fermion
produced in the tree-level decay $\fdm\to\ell^+\ell^-\nn$ is comparable in
size to the dark matter mass itself. This possibility can occur, for example,
within the leptophilic model discussed in section~\ref{sec:Examples}, where
$\fdm$ is the hidden gaugino of an unbroken $U(1)$-symmetry, and $\nn$ is a
neutralino~\cite{Ibarra:2008kn}.

As was shown in section~\ref{sec:scalar_intensity}, the decay channel
$\fdm\to\gamma\nn$ is kinematically enhanced compared to the three-body decay
when $m_\nn \sim m_\fdm$, provided that $\fdm$ and $\nn$ have opposite
\textsl{CP} parities ($\eta=-1$). Thus, such scenarios can be tested
particularly well via the loop-induced gamma-ray line signal. In order to
infer the observational constraints, it is convenient to consider the ratio
$m_\fdm/\Gamma(\psi_\text{DM}\to\gamma N)$, which determines the magnitude of
the observable flux. For the case of scalar-mediated decay, and purely chiral
couplings, it is given by (see eq.~\eqref{ratioScalar})
\begin{eqnarray}
  \lefteqn{ \Gamma(\fdm\to\gamma\nn)^{-1}m_\fdm \ \approx \
  \frac{1}{R^\text{chir}} \left( \frac{\Gamma(\fdm\to\ell^+\ell^-)^{-1}m_\fdm
  }{10^{26}\sec \times 2.5\TeV } \right) \times } \nonumber\\ && {} \times
  \left\{\begin{array}{ll} 3\times 10^{29}\sec\TeV\, & \mbox{for } m_\nn \to
  0, ~ \eta = \pm 1\\ 7\times 10^{29}\sec\TeV\, & \mbox{for } m_\nn \to
  m_\fdm, ~ \eta = +1\\ 1.4\times 10^{28}\sec\TeV\,
  \left(\frac{2E_\gamma}{m_\fdm}\right)^2 & \mbox{for } m_\nn \to m_\fdm, ~
  \eta = -1 \end{array}\right. \;,
\end{eqnarray}
where $R^\text{chir}=1$ for three-body decays into a single lepton flavor, and
$R^\text{chir}=3$ for flavor-democratic three-body decays. In the case of an
intermediate vector, the right-hand side is smaller by a factor nine, implying
a nine times larger gamma-ray flux.  From the last line, it is apparent that
in the case of opposite \textsl{CP} parities, the monochromatic gamma-ray flux
is enhanced for large values of $m_\fdm$, when keeping the photon energy
$E_\gamma$ fixed.  Note that a similar enhancement of the decay channel
$\fdm\to Z^0\nn$, which may also be induced at the loop-level, could lead to
complementary constraints from the antiproton flux produced by the
fragmentation of the $Z$-boson, which we do not discuss here.

In order to illustrate this result, we consider a number of benchmark
scenarios for which $m_\fdm$ and $m_\nn$ are of comparable size, with
parameters chosen as shown in table~\ref{tab:benchmarks}. All the benchmark
scenarios reproduce the PAMELA positron data, and all except scenarios 1, 2
and 3 additionally reproduce the electron spectrum measured by Fermi. Note that
the maximum lepton energy in the three-body decay $\fdm\to\ell^+\ell^-\nn$
coincides with the energy of the monochromatic photons,
$E_\text{max}=E_\gamma$.

The gamma-ray line signal induced by the one-loop decay $\fdm\to\gamma\nn$ is
shown in fig.~\ref{fig:bounds_benchmarks} for the various benchmark scenarios.
Clearly, the scenarios 1 and 4 are in conflict with the gamma-ray line
searches performed by Fermi and HEGRA, respectively. Thus, despite the fact
that the dark matter couples only to leptons at tree-level, the gamma-ray line
signal induced by one-loop corrections has an intensity that is detectable by
present gamma-ray telescopes. In other words, scenarios 1 and 4 can be ruled
out as possible explanations of the high-energy positron excess, because the
loop-induced radiative decay produces a gamma-ray line that should have been
already detected. This shows that the higher-order corrections are indeed
relevant and have to be taken into account.  In contrast, the other benchmarks
are in agreement with present bounds on gamma-ray lines. For example, in
scenario 3 the partial lifetime for the radiative decay is larger compared to
scenario 1, and lies slightly above the current Fermi bounds. Scenario 6 can
be tested in the future by the CTA. Since there is no kinematic enhancement
of the decay $\fdm\to\gamma\nn$ in the case $\eta=+1$, the intensity of the
gamma-ray line is comparably weak in scenarios 2 and 5.  For example, scenario
5 differs from 6 just by the sign of $\eta$, but is much more difficult to
probe by the CTA. Finally, for benchmark point 7, we assume that the couplings
of the leptons to $\fdm$ and to $\nn$ have opposite chirality, in which case
the loop is strongly suppressed and there is no hope of detecting a gamma-ray
line signal.

\section{Conclusions}\label{sec:Conclusions}
We have analyzed the radiative decay of dark matter particles in view of the
leptonic cosmic-ray anomalies reported by PAMELA and Fermi LAT. Assuming an
effective description of leptophilic dark matter decay, we have pointed out
that the lines induced at the quantum level may be observable and can be used
to constrain models of decaying dark matter. In the case of scalar dark
matter, two-body decays into photons are strongly helicity-suppressed and thus
unobservable. In the case that the dark matter particles carry spin 1/2,
however, the radiative decay rate is typically suppressed compared to the
tree-level decays by some two to three orders of magnitude. Interestingly, the
corresponding partial lifetimes for decays into monochromatic photons can then
be in the observable range, in particular for dark matter masses of a few
hundred GeV, where stringent constraints from Fermi LAT apply. Thus, in some
cases the loop-induced gamma-ray line yields constraints that can be
competitive with the constraints on charged cosmic rays. At higher energies,
constraints from Cherenkov telescopes exist. At present, these constraints are
only relevant for certain scenarios for which the radiative two-body decay is
kinematically enhanced compared to the three-body decay channel.  However, we
have pointed out that the proposed CTA should be able to improve on the
existing bounds significantly and probe a relevant part of the parameter space
which is presently unconstrained.

\section*{Acknowledgements}
CW is grateful to Pierre Colin, Dieter Horns and Daniela Borla Tridon for
valuable discussions. The work of MG, AI and DT was supported by the DFG
cluster of excellence ``Origin and Structure of the Universe.'' DT also
acknowledges support from the DFG Graduiertenkolleg ``Particle Physics at the
Energy Frontier of New Phenomena.''

\appendix

\section{Decay widths for fermionic dark matter}
\label{app:fermion}
\subsection{The decay $\psi_\text{DM} \rightarrow \ell^+ \ell^- N$}
The differential decay rate for this process is given by
\begin{equation}
  d\Gamma(\psi_\text{DM} \rightarrow \ell^+ \ell^- N) = \frac{1}{(2 \pi)^3}
  \frac{1}{64 m_{\psi_\text{DM}}^3} |\mathcal{M}_t + \mathcal{M}_u|^2 dt \, ds
  \;.
\end{equation}
Note that there is a relative minus sign between the $t$- and $u$-channel
amplitudes due to the exchange of two anticommuting fermions that is not
present by a naive application of the Feynman rules for the two diagrams.
Neglecting the lepton mass, one obtains for the squared amplitude
\begin{align}
  |\mathcal{M}_t + \mathcal{M}_u|^2 = \, &\left(|\lambda_{\ell \psi}^L|^2 +
  |\lambda_{\ell \psi}^R|^2\right) \left(|\lambda_{\ell N}^L|^2 +
  |\lambda_{\ell N}^R|^2\right) \nonumber\\ & \times \left[\frac{(t -
  m_N^2)(m_{\psi_\text{DM}}^2 - t)}{(t - m_\Sigma^2)^2} + \frac{(u -
  m_N^2)(m_{\psi_\text{DM}}^2 - u)}{(u - m_\Sigma^2)^2}\right] \nonumber\\ &+
  2 \eta \Bigg\{\text{Re}\left[\left(\lambda_{\ell \psi}^{L*} \lambda_{\ell
  N}^L\right)^2 + \left(\lambda_{\ell \psi}^{R*} \lambda_{\ell
  N}^R\right)^2\right] \frac{m_{\psi_\text{DM}} m_N s}{(t - m_\Sigma^2)(u -
  m_\Sigma^2)} \nonumber\\ &- \text{Re}\left[\lambda_{\ell \psi}^{L*}
  \lambda_{\ell N}^L \lambda_{\ell \psi}^{R*} \lambda_{\ell N}^R\right] \times
  \nonumber\\ &\times \frac{(t - m_N^2)(m_{\psi_\text{DM}}^2 - t) + (u -
  m_N^2)(m_{\psi_\text{DM}}^2 - u) - s(t + u)}{(t - m_\Sigma^2)(u -
  m_\Sigma^2)}\;,
\end{align}
where
\begin{equation}
  s = (q_1 - p_1)^2, ~~~ t = (q_1 - p_2)^2, ~~~ u = (q_1 - p_3)^2 = m_\psi^2 +
  m_N^2 + 2 m_\ell^2 - s - t\;.
\end{equation}
Again, $\eta = \eta_{\psi_\text{DM}} \eta_N = \pm 1$ depending on the
\textsl{CP} eigenvalues of $\psi_\text{DM}$ and $N$. The integration limits
for the Mandelstam variables are given by
\begin{equation}
  0 \leq s \leq (m_{\psi_\text{DM}} - m_N)^2
\end{equation}
and
\begin{equation}
  t_{1,2} = \frac{1}{2} \left(m_{\psi_\text{DM}}^2 + m_N^2 - s \mp
  \sqrt{\lambda(m_{\psi_\text{DM}}^2,m_N^2,s)}\right)\;,
\end{equation}
where
\begin{equation}
  \lambda(a,b,c) = a^2 + b^2 + c^2 - 2 ab - 2 ac - 2 bc\;.
\end{equation}
We can perform the kinematical integrations in the limit $m_\Sigma \gg t,u$,
in which case the Mandelstam variables in the denominator can be neglected. We
then get
\begin{align}
  \Gamma(\psi_\text{DM} \rightarrow \ell^+ \ell^- N) = {} & \frac{1}{64 (2
  \pi)^3} \frac{m_{\psi_\text{DM}}^5}{6 m_\Sigma^4}
  \Bigg\{\Big[\left(|\lambda_{\ell \psi}^L|^2 + |\lambda_{\ell
  \psi}^R|^2\right) \left(|\lambda_{\ell N}^L|^2 + |\lambda_{\ell
  N}^R|^2\right) \nonumber\\ & - \eta \, \text{Re} \left(\lambda_{\ell
  \psi}^{L*} \lambda_{\ell N}^L \lambda_{\ell \psi}^{R*} \lambda_{\ell
  N}^R\right) \Big] F_1(x)\nonumber\\ & + 2 \eta \, \text{Re}
  \left[\left(\lambda_{\ell \psi}^{L*} \lambda_{\ell N}^L\right)^2 +
  \left(\lambda_{\ell \psi}^{R*} \lambda_{\ell N}^R\right)^2\right] F_2(x)
  \Bigg\}\;,
\end{align}
where $x \equiv m_N^2/m_{\psi_\text{DM}}^2$ and $F_1(x)$, $F_2(x)$ are defined
in eqs. (\ref{eqn:F1}), (\ref{eqn:F2}).

In the case of mediation by a vector, the matrix element for vanishing lepton
mass reads
\begin{align}
  |\mathcal{M}_t + \mathcal{M}_u|^2 = & ~ 4 \left(\left|\lambda_{\ell \psi}^L
  \lambda_{\ell N}^L\right|^2 + \left|\lambda_{\ell \psi}^R \lambda_{\ell
  N}^R\right|^2\right) \nonumber\\ & \times \left[\frac{(u -
  m_N^2)(m_{\psi_\text{DM}}^2 - u)}{(t - m_V^2)^2} + \frac{(t -
  m_N^2)(m_{\psi_\text{DM}}^2 - t)}{(u - m_V^2)^2}\right] \nonumber\\ & + 4
  \left(\left|\lambda_{\ell \psi}^L \lambda_{\ell N}^R\right|^2 +
  \left|\lambda_{\ell \psi}^R \lambda_{\ell N}^L\right|^2\right) s(t + u)
  \left[\frac{1}{(t - m_V^2)^2} + \frac{1}{(u - m_V^2)^2}\right] \nonumber\\ &
  + 8 \eta \, \Bigg\{\text{Re}\left[\left(\lambda_{\ell
  \psi}^L \lambda_{\ell N}^{L*}\right)^2 + \left(\lambda_{\ell \psi}^R
  \lambda_{\ell N}^{R*}\right)^2\right] \frac{m_{\psi_\text{DM}} m_N s}{(t -
  m_V^2) (u - m_V^2)} \nonumber\\ & + 2 \text{Re} \left[\lambda_{\ell \psi}^L
  \lambda_{\ell N}^{L*} \lambda_{\ell \psi}^R \lambda_{\ell N}^{R*}\right]
  \frac{s(t + u)}{(t - m_V^2)(u - m_V^2)}\Bigg\}\;.
\end{align}
In the limit $m_{\psi_\text{DM}} \ll m_V$ we get for the decay rate
\begin{align}
  \Gamma(\psi_\text{DM} \rightarrow \ell^+ \ell^- N) = {} & \frac{1}{64
  (2\pi)^3} \frac{4 m_{\psi_\text{DM}}^5}{6 m_V^4}
  \Bigg\{\Big[\left(|\lambda_{\ell \psi}^L|^2 + |\lambda_{\ell
  \psi}^R|^2\right) \left(|\lambda_{\ell N}^L|^2 + |\lambda_{\ell
  N}^R|^2\right) \nonumber\\ & {} + 2\eta \,
  \text{Re}\left(\lambda_{\ell \psi}^L \lambda_{\ell N}^{L*} \lambda_{\ell
  \psi}^R \lambda_{\ell N}^{R*}\right)\Big] F_1(x) \nonumber\\ & {} +
  2\eta \, \text{Re}\left[\left(\lambda_{\ell \psi}^L
  \lambda_{\ell N}^{L*}\right)^2 + \left(\lambda_{\ell \psi}^R \lambda_{\ell
  N}^{R*}\right)^2\right] F_2(x)\Bigg\}\;.
\end{align}

\subsection{The decay $\psi_\text{DM} \rightarrow \gamma N$}
There are four scalar-mediated diagrams at the one-loop level contributing to
the decay $\psi_\text{DM} \rightarrow \gamma N$, which are shown in
fig.~\ref{fermion_loop}. Due to gauge invariance, in the case of
\textsl{CP}-conserving interactions, the matrix element corresponding to the
sum of the four diagrams can be written in the form
\begin{align}
  \nonumber \mathcal{M} &= \frac{i g_{N \gamma \psi}}{m_{\psi_\text{DM}}}
  \bar{u}(k_1)(P_R - \eta_N \eta_\psi P_L) \sigma^{\mu \nu} k_{2 \mu}
  \epsilon_\nu^* u(p)\\ &= -\frac{g_{N \gamma \psi}}{m_{\psi_\text{DM}}}
  \bar{u}(k_1)(P_R - \eta_N \eta_\psi P_L) \slashed{k}_2 \slashed{\epsilon}^*
  u(p)\;,
\end{align}
where $\sigma^{\mu \nu} = i [\gamma^\mu,\gamma^\nu]/2$ and
$\eta_{\psi_\text{DM}}$, $\eta_N$ are the \textsl{CP} eigenvalues of
$\psi_\text{DM}$ and $N$, respectively. This is manifestly gauge invariant in
the sense that it satisfies the Ward identity: the matrix element vanishes
when replacing $\epsilon_\mu^* \rightarrow k_{2 \mu}$ since the photon is
on-shell.
 
The effective coupling $g_{N \gamma \psi_\text{DM}}^\Sigma$ for an
intermediate scalar can be given in terms of loop integrals as follows,
\begin{align}
  \nonumber
  g_{N \gamma \psi_\text{DM}}^\Sigma = &-\frac{e \, \eta_N
  m_{\psi_\text{DM}}}{16 \pi^2} \sum_{f,\Sigma} Q_f C_f \Big\{m_f
  (\eta_{\psi_\text{DM}} \lambda_{\ell N}^L \lambda_{\ell \psi}^R - \eta_N
  \lambda_{\ell N}^R \lambda_{\ell \psi}^L) I\\ &+ (\lambda_{\ell N}^L
  \lambda_{\ell \psi}^L - \eta_{\psi_\text{DM}} \eta_N \lambda_{\ell N}^R
  \lambda_{\ell \psi}^R) [\eta_{\psi_\text{DM}} m_{\psi_\text{DM}} (I^2 - K) -
  \eta_N m_N K]\Big\}\;,
\end{align}
where the sum runs over all fermions $f$ and all mediators $\Sigma$ that
contribute in the loop. The loop integrals are written in terms of Feynman
parameters as
\begin{align}
  I &= \frac{1}{\Delta} \int_0^1 \frac{dx}{1 - x} \log X\\ I^2 &=
  \frac{1}{\Delta} \int_0^1 dx \log X\\ K &= -\frac{1}{\Delta} \int_0^1 dx
  \left(1 + \frac{B}{\Delta \, x(1 - x)} \log X\right)\;,
\end{align}
where
\begin{align}
  \Delta &\equiv m_{\psi_\text{DM}}^2 - m_N^2\\ B &\equiv m_\ell^2 x +
  m_\Sigma^2 (1 - x) - m_{\psi_\text{DM}}^2 x (1 - x)\\ X &\equiv
  \frac{m_\ell^2 x + m_\Sigma^2 (1 - x) - m_{\psi_\text{DM}}^2 x (1 -
  x)}{m_\ell^2 x + m_\Sigma^2 (1 - x) - m_N^2 x (1 - x)}\;.
\end{align}
In the limit $m_{\psi_\text{DM}},m_N \ll m_\Sigma$, the loop integrals take on
the simplified form~\cite{Haber:1988px}
\begin{align}
  I &= \frac{1}{m_\Sigma^2} f(m_\ell^2/m_\Sigma^2),\\
  I^2 &= -\frac{1}{2 m_\Sigma^2} f_2(m_\ell^2/m_\Sigma^2),\\
  K &= \frac{1}{2} I^2,
\end{align}
where the functions $f$, $f_2$ are defined as
\begin{align}
  f(x) &= \frac{1}{1 - x} \left[1 + \frac{1}{1 - x} \ln(x)\right],\\
  f_2(x) &= \frac{1}{(1 - x)^2} \left[1 + x + \frac{2x}{1 - x} \ln(x)\right].
\end{align}
The expression for the effective coupling then assumes the form
\begin{align}\label{eqn:fermion_g_simplified}
  g_{N \gamma \psi}^\Sigma \simeq & -\frac{e \, \eta_N m_{\psi_\text{DM}}}{16
  \pi^2} \sum_{\ell,\Sigma} Q_\ell C_\ell \Big\{m_\ell
  \left(\eta_{\psi_\text{DM}} \lambda_{\ell N}^L \lambda_{\ell \psi}^R -
  \eta_N \lambda_{\ell N}^R \lambda_{\ell \psi}^L\right)
  \frac{f(m_\ell^2/m_\Sigma^2)}{m_\Sigma^2} \nonumber\\\nonumber & {} -
  \left(\lambda_{\ell N}^L \lambda_{\ell \psi}^L - \eta_{\psi_\text{DM}}
  \eta_N \lambda_{\ell N}^R \lambda_{\ell \psi}^R\right)
  \frac{\eta_{\psi_\text{DM}} m_{\psi_\text{DM}} - \eta_N m_N}{4 m_\Sigma^2}
  f_2(m_\ell^2/m_\Sigma^2)\Big\}\\ {} \simeq & \, \frac{e \, \eta}{64\pi^2}
  m_{\psi_\text{DM}}^2 \left(1 - \frac{\eta \, m_N}{m_{\psi_\text{DM}}}\right)
  \sum_{\ell,\Sigma} \frac{Q_\ell
  C_\ell}{m_\Sigma^2}\left\{\left(\lambda_{\ell N}^L \lambda_{\ell \psi}^L -
  \eta \, \lambda_{\ell N}^R \lambda_{\ell \psi}^R\right)\right\}\;,
\end{align}
where in the last line we have taken $m_\ell \rightarrow 0$.

In the case of an intermediate vector, the same integrals $I$, $I^2$, $K$
(with the replacement $m_\Sigma \rightarrow m_V$ in the constants $B$, $X$)
appear, together with the additional integral
\begin{equation}
  J = \frac{1}{\Delta} \int_0^1 \frac{dx}{x} \log X\;,
\end{equation}
which simplifies in the limit $m_{\psi_\text{DM}},m_N \ll m_\Sigma$ to
\begin{align}
  J &= \frac{1}{m_V^2} \frac{\ln(x)}{1 - x} - I = -\frac{1}{m_V^2}
  f^V(m_\ell^2/m_V^2)\;,
\end{align}
where in this case the kinematical functions are defined as
\begin{align}
  f^V(x) &= \frac{1}{1 - x} \left[1 + \frac{x}{1 - x} \ln(x)\right]\\ f_2^V
  (x) &= \frac{1}{(1 - x)^2} \left[1 - \frac{5x}{3} + \frac{2x (1 - 2x)}{3 (1
  - x)} \ln(x)\right]\;.
\end{align}
Furthermore, one finds
\begin{equation}
  I^2 - J - K = K - J = \frac{3}{4 m_V^2} f^V(x)\;.
\end{equation}
The effective coupling in terms of loop integrals is given by
\begin{align}
  g_{N \gamma \psi}^V = {}& \frac{e \, \eta_N m_{\psi_\text{DM}}}{8 \pi^2}
  \sum_{\ell} \Big\{(\eta \lambda_{\ell N}^L \lambda_{\ell
  \psi}^L - \lambda_{\ell N}^R \lambda_{\ell \psi}^R)
  \big[\eta_{\psi_\text{DM}} m_{\psi_\text{DM}} (I^2 - J - K) \nonumber\\ & +
  \eta_N m_N (J - K)\big] + 2 m_\ell (\eta_{\psi_\text{DM}} \lambda_{\ell N}^L
  \lambda_{\ell \psi}^R - \eta_N \lambda_{\ell N}^R \lambda_{\ell \psi}^L)
  J\Big\}\;,
\end{align}
For $m_{\psi_\text{DM}} \ll m_V$ this expression then simplifies to
\begin{align}
  g_{N \gamma \psi}^V \simeq & -\frac{e \, \eta_N m_{\psi_\text{DM}}}{8 \pi^2}
  \sum_\ell \Big\{2 m_\ell \left(\eta_N \lambda_{\ell N}^L \lambda_{\ell
  \psi}^R - \eta_{\psi_\text{DM}} \lambda_{\ell N}^R \lambda_{\ell
  \psi}^L\right) \frac{f^V(m_\ell/m_V^2)}{m_V^2} \nonumber\\\nonumber & - 3
  \left(\eta_N \eta_{\psi_\text{DM}} \lambda_{\ell N}^L \lambda_{\ell \psi}^L
  - \lambda_{\ell N}^R \lambda_{\ell \psi}^R\right)
  \frac{\eta_{\psi_\text{DM}} m_{\psi_\text{DM}} - \eta_N m_N}{4 m_V^4}
  f_2^V(m_\ell^2/m_V^2) \Big\}\\ \simeq {} & \frac{3e \, \eta}{32\pi^2}
  \frac{m_{\psi_\text{DM}}^2}{m_V^2} \left(1 - \frac{\eta \,
  m_N}{m_{\psi_\text{DM}}}\right) \sum_\ell \left(\lambda_{\ell N}^L
  \lambda_{\ell \psi}^L - \eta \, \lambda_{\ell N}^R \lambda_{\ell
  \psi}^R\right)
\end{align}
where in the last line we have taken $m_\ell \rightarrow 0$.

The decay rate in both cases is finally given by
\begin{equation}
  \Gamma(\psi_\text{DM} \rightarrow \gamma N) = \frac{\left(g_{N \gamma
  \psi}^{\Sigma/V}\right)^2}{8 \pi} m_{\psi_\text{DM}} \left(1 -
  \frac{m_N^2}{m_{\psi_\text{DM}}^2}\right)^3.
\end{equation}
For the scalar in the case of one mediator coupled to leptons we get in the
limit $m_\ell \ll m_N$ and $m_{\psi_\text{DM}} \ll m_\Sigma$,
\begin{align}
  \Gamma(\psi_\text{DM} \rightarrow \gamma N) = {}& \frac{e^2}{8 \pi
  \left(64\pi^2\right)^2} \frac{m_{\psi_\text{DM}}^5}{m_\Sigma^4} \left(1 -
  \frac{m_N^2}{m_{\psi_\text{DM}}^2}\right)^3 \left(1 - \frac{\eta \,
  m_N}{m_{\psi_\text{DM}}}\right)^2 \nonumber\\ & \times \left[\sum_\ell
  \left(\lambda_{\ell N}^L \lambda_{\ell \psi}^L - \eta \, \lambda_{\ell N}^R
  \lambda_{\ell \psi}^R\right)\right]^2\;,
\end{align}
whereas for the vector we have, in the limit $m_\ell \ll m_N$ and
$m_{\psi_\text{DM}} \ll m_V$,
\begin{align}
  \Gamma(\psi_\text{DM} \rightarrow \gamma N) &= \frac{1}{8\pi}
  \frac{9e^2}{\left(8\pi^2\right)^2} \frac{m_{\psi_\text{DM}}^5}{16 m_V^4}
  \left(1 - \frac{m_N^2}{m_{\psi_\text{DM}}^2}\right)^3 \left(1 - \frac{\eta
  \, m_N}{m_{\psi_\text{DM}}}\right)^2 \nonumber\\ & ~~~ \times
  \left[\sum_\ell \left(\lambda_{\ell N}^L \lambda_{\ell \psi}^L - \eta \,
  \lambda_{\ell N}^R \lambda_{\ell \psi}^R\right)\right]^2\;.
\end{align}

\section{Decay widths for scalar dark matter}
\label{app:scalar}

In this appendix we present the expressions for the decay width of the
radiative decay of scalar dark matter into two photons, $\phi_\text{DM}
\rightarrow \gamma \gamma$.

For $\lambda_{\ell \phi}^L = \lambda_{\ell \phi}^R \equiv \lambda_{\ell \phi}$
the decay rate reads \cite{Resnick:1973vg,Spira:1995rr}
\begin{equation}
  \Gamma(\phi_\text{DM} \rightarrow \gamma \gamma) =
  \frac{m_{\phi_\text{DM}}^3}{4 \pi} \left(\frac{e^2}{16 \pi^2}\right)^2
  \left|\sum_\ell \frac{\lambda_{\ell \phi}}{m_\ell} A_f(\tau_\ell)\right|^2,
\end{equation}
where $\tau_\ell \equiv m_{\phi_\text{DM}}^2/(4 m_\ell^2)$ and
\begin{align}
  A_f(\tau) &= 2 \left[\tau + (\tau - 1) f(\tau)\right] / \tau^2\\ f(\tau) &=
  \begin{cases} \arcsin^2 \sqrt{\tau} \, , &\tau \leq 1\\ -\frac{1}{4}
  \left[\ln \frac{1 + \sqrt{1 - 1/\tau}}{1 - \sqrt{1 - 1/\tau}} - i
  \pi\right]^2, &\tau > 1 \label{eqn:ftau} \end{cases}\;.
\end{align}
In the case of interest here, $\tau_\ell \gg 1$. Then we can approximate
\begin{equation}
  A_f(\tau) \simeq \frac{1}{\tau} \left\{2 - \frac{1}{2} (\ln(4 \tau) - i
  \pi)^2\right\} \;.
\end{equation}
In this limit, and taking only one lepton species into account, the decay rate
is given by
\begin{align}
  \Gamma(\phi_\text{DM} \rightarrow \gamma \gamma) \simeq {} &
  \frac{|\lambda_{\ell \phi}|^2}{16 \pi} m_{\phi_\text{DM}}
  \left(\frac{e^2}{16 \pi^2}\right)^2 \frac{4 m_\ell^2}{m_{\phi_\text{DM}}^2}
  \nonumber\\
  & \times \left\{\left[2 + \frac{\pi^2}{2} - \frac{1}{2} \ln^2(4
  \tau_\ell)\right]^2 + \pi^2 \ln^2(4 \tau_\ell)\right\}\;.
\end{align}

\end{document}